# CVD Grown Hybrid MoSe$_2$-WSe$_2$ Lateral/Vertical Heterostructures with Strong Interlayer Exciton Emission

Md Tarik Hossain[1]*, Sai Shradha[2], Axel Printschler[1], Julian Picker[1], Luc F. Oswald[2], Julian Führer[2], Nicole Engel[2], Honey Jayeshkumar Shah[1], Christof Neumann[1], Daria I. Markina[2], Moritz Quincke[3], Johannes Biskupek[3], Kenji Watanabe[4], Takashi Taniguchi[5], Ute Kaiser[3], Bernhard Urbaszek[2], Andrey Turchanin[1,6]*

[1]Institute of Physical Chemistry, Friedrich Schiller University Jena,
Lessingstr. 10, 07743 Jena, Germany

[2]Institute of Condensed Matter Physics, Technical University Darmstadt,
Hochschulstraße 6-8, 64289 Darmstadt, Germany

[3]Central Facility of Electron Microscopy, Electron Microscopy Group of Material Science, Ulm University, 89081 Ulm, Germany

[4]Research Center for Electronic and Optical Materials,
National Institute for Materials Science, 1-1 Namiki, Tsukuba 305-0044, Japan

[5]Research Center for Materials Nanoarchitectonics,
National Institute for Materials Science, 1-1 Namiki, Tsukuba 305-0044, Japan

[6]Abbe Center of Photonics, Albert-Einstein-Straße 6, 07745 Jena, Germany

**Corresponding Authors**:

Md Tarik Hossain (tarik.hossain@uni-jena.de),

Andrey Turchanin (andrey.turchanin@uni-jena.de)

**ABSTRACT**

Lateral heterostructures of 2D transition metal dichalcogenide offer a powerful platform to investigate photonic and electronic phenomena at atomically sharp interfaces. However, their controlled engineering, including tuning lateral domain size and integration into vertical van der Waals heterostructures with other 2D materials, remains challenging. Here, we present a facile route for the synthesis of two types of heterostructures consisting of monolayers of $MoSe_2$ and $WSe_2$ - purely lateral (HS I) and hybrid lateral/vertical (HS II) - using liquid precursors of transition metal salts and chemical vapor deposition (CVD). Depending on the growth parameters, the heterostructure type and their lateral dimensions can be adjusted. We characterized properties of the HS I and HS II by complementary spectroscopic and microscopic techniques including Raman and photoluminescence spectroscopy, and optical and atomic force microscopy, and scanning electron and transmission electron microscopy. The photoluminescence measurements reveal strong interlayer exciton emission in the $MoSe_2$/$WSe_2$ region of HS II, which dominates the spectrum at 4 K and persisting up to room temperature. These results demonstrate high optical quality of the grown heterostructures which in combination with scalability of the developed approach paves the way for fundamental studies and device applications based on these unique 2D quantum materials.

# 1 Introduction

Due to their atomic thickness and unique electronic and photonic properties, heterostructures (HSs) of monolayer (ML) transition metal dichalcogenides (TMDs), in both lateral and vertical configurations, hold significant promise for novel electronic, optoelectronic and photonic devices as well as for studying new quantum phenomena.[1, 2] They enable investigations of new solid-state phenomena like interlayer excitons[3] and charge transfer excitons[4], electric transport through 1D *p-n* junctions[5], which are not accessible in individual TMD MLs. Based on the similarity of the crystal structures, 2D TMD can be stitched or stacked to form $TMD_1$-$TMD_2$ lateral junction or vertical van der Waals $TMD_1$/$TMD_2$ HSs. Both HSs offer new opportunities for photonics hosting various excitonic phenomena and enabling their device integration (see e.g. [3, 6-8]). For instance, lateral HSs demonstrate unidirectional in-plane exciton transport[9], which facilitates development of excitonic devices. The spatial modulation of electronic properties in LHs enables the formation of atomically thin in-plane *p-n* junctions for the engineering of transistors, diodes, photodetectors, photovoltaic devices, electroluminescent devices, and other quantum devices.[5, 10-13] On the other hand, vertical HSs demonstrate the formation of interlayer excitons (IEs), where electrons and holes are separated between the two adjacent TMD MLs (see e.g. [3, 14-17]). As an example, $MoSe_2$/$WSe_2$ vertical HS form the type-II band alignment resulting in a rapid interlayer charge transfer between $MoSe_2$ and $WSe_2$ and in the formation of IEs with long lifetimes,[3, 18, 19] which facilitates development of fields such as quantum encryption and valleytronics[20-23]. Furthermore, Moiré trapped IEs, owing to their discrete emission, may be used as single photon emitters[23] and for lasing via coupling to nanocavities[24].

For fundamental studies and applications, engineering of TMD HSs by bottom-up approaches is particularly important as it enables scalability and integration into semiconductor and optical technologies. Typically, either one-step[7, 25-31] or two/multi-step[13, 32-34] chemical vapor deposition (CVD) growth methods using powders of transition metal oxides (e.g., $MoO_3$, $WO_3$) and chalcogens (S, Se) or a mixture of transition metal dichalcogenide (e.g., $WSe_2$ and $MoSe_2$) as precursors are utilized to this end.[2, 35] In a more conventional two-step growth method, MLs of two different TMDs are grown sequentially employing separate growth reactors.[32] Such a method poses challenges, as the exposure of the first TMD to the ambient conditions before the growth of the second one can compromise the quality of the atomic interfaces. Alternatively, a one-step CVD enables the growth within the same reactor without exposing the sample to ambient conditions. However, the use of powder metal oxides limits the ability to tune the HS sizes due to difficulties in controlling their vapor pressures and fluxes of the precursors during the growth. On the other hand, water soluble sodium molybdate ($Na_2MoO_4$) and sodium tungstate ($Na_2WO_4$) salts provide flexibility in tuning the transition metal ratios in the precursors by their appropriate mixing and therefore possess great promise to overcome this limitation. To the best of our knowledge, up to date only a few studies reported CVD growth of HSs using liquid precursors.[36-39] For instance, metal-organic liquid precursors have been employed to grow $TMD_1$-$TMD_2$ lateral HSs[36], while ammonium- and sodium-based salts have been used for the growth of $MoS_2$-$WS_2$ lateral[39] and vertical HSs to investigate their electrical transport properties[37]. Similarly, sodium-based salts have enabled the synthesis of $MoSe_2$-$WSe_2$ lateral HSs for photodetector applications[38]. Nevertheless, the realization of both vertical and lateral HS with clean and atomically sharp interfaces for studying excitonic phenomena remains a challenge and in this respect bottom-up approaches hold a significantly higher potential in comparison to their top-down assembly from individual MLs, where contamination is intrinsically unavoidable.

Here, we demonstrate a tunable, highly reproducible and scalable one-step CVD approach, utilizing liquid precursors of transition metals, to grow purely lateral (HS I) as well as hybrid lateral/vertical (HS II) HSs of MoSe$_2$ and WSe$_2$ MLs, *Figure 1*. To obtain the desired HSs, aqueous solutions of Na$_2$MoO$_4$ and Na$_2$WO$_4$ salts with different concentrations were spin-coated onto 300 nm SiO$_2$/Si growth substrates, followed by a CVD conversion step at elevated temperatures in the presence of Se vapor. The formation of HS I and HS II, their sizes and ratios of MoSe$_2$ and WSe$_2$ MLs' areas can be tuned by the concentration of the salts and by the growth temperature. We employed complementary microscopic and spectroscopic techniques, including optical microscopy (OM), atomic force microscopy (AFM), Raman spectroscopy, photoluminescence (PL) spectroscopy, and transmission electron microscopy (TEM) including high-resolution (HR) and scanning (STEM) together with energy dispersive X-ray spectroscopy (EDX) to study their morphological, structural, chemical, and optical properties. By PL measurements, we reveal over a broad temperatures range (from 4 K to 300 K) a strong IE emission in the WSe$_2$/MoSe$_2$ region of HS II. The room temperature (RT) IE formation is remarkable and demonstrates an efficient carrier transfer between the CVD grown WSe$_2$ and MoSe$_2$ MLs. It is also striking that the IE at T=4 K dominates the PL spectrum, which has only been observed in the HSs obtained by mechanical stacking of individual TMD MLs exfoliated from bulk crystals[40] or grown via the CVD method from solid precursors[41, 42].

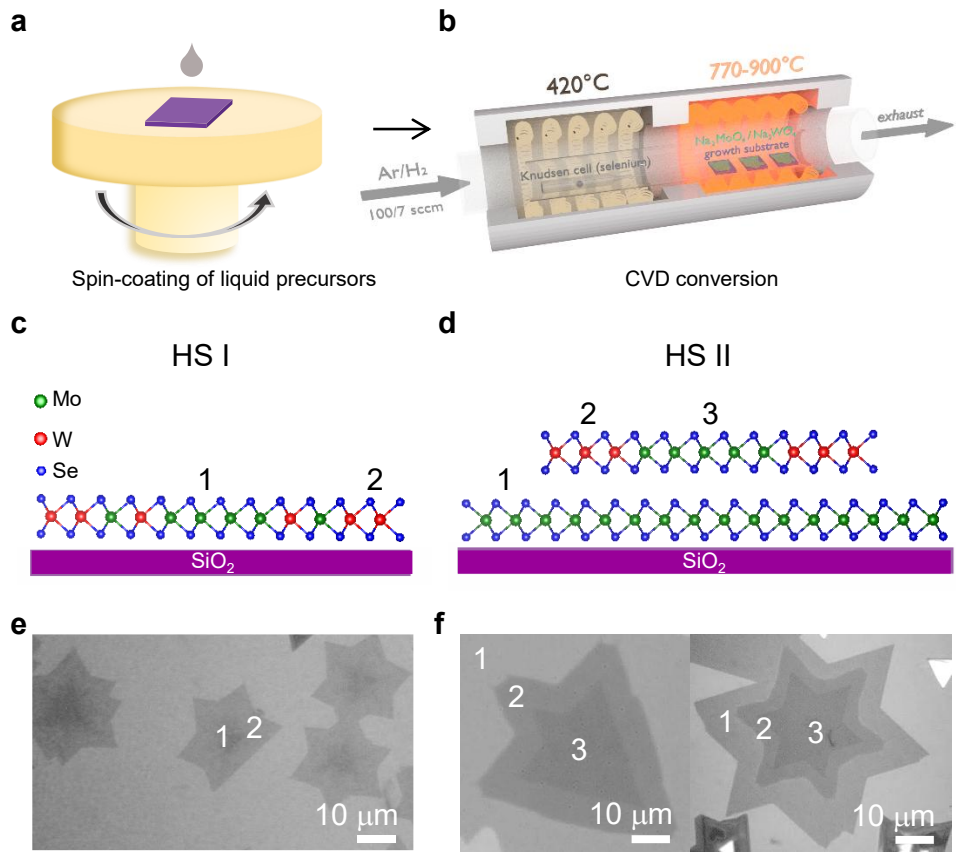

**Figure 1:** Growth steps and structural configurations of $MoSe_2$-$WSe_2$ heterostructures. (a-b) Schematics illustrating the growth process for $MoSe_2$ and $WSe_2$ heterostructures, which includes the preparation of precursor substrate and subsequent CVD conversion. (c-d) Schematics of two heterostructure configurations: HS I, where $MoSe_2$ and $WSe_2$ stitched laterally, and HS II, where a $MoSe_2$ bottom layer is vertically stacked with a $MoSe_2$-$WSe_2$ lateral heterostructure grown on top. (e-f) The representative optical microscopy images of HS I and HS II, respectively.

## 2 Results and Discussion

### 2.1 CVD growth of HS I and HS II using liquid precursors

We begin with a general description of the preparation method of HS I and HS II illustrated in *Figure 1a,b*. To this end, the liquid $Na_2MoO_4$ and $Na_2WO_4$ precursors, which served as the molybdenum (Mo) and tungsten (W) sources, respectively, were deposited onto a 300 nm $SiO_2$/Si growth substrate and subsequently converted into the HSs in a CVD reactor where solid selenium powder (Se) was used as chalcogen precursor. By adjusting the Mo:W precursors molar ratios (1:3 or 3:1; Table S1, SI), two distinct configurations were obtained. HS I consists of a $MoSe_2$ core laterally surrounded by $WSe_2$, forming hexagonal, star-like, or triangular crystals (*Figure 1c, e, and Figure S1*). HS II exhibits a vertical bilayer structure with a $MoSe_2$ base and a lateral $MoSe_2$-$WSe_2$ top layer with triangle or hexagonal edges (*Figure 1d, f, and Figure S2*). For growth, a modified CVD reactor equipped with quartz Knudsen-type effusion cells was employed, thereby enabling precise control of the selenium vapor flux [43]. The reactor furnace was heated to growth temperatures of 770, 820, or 900 °C at a rate of 15 °C min$^{-1}$ (*Figure S3*). Given the lower melting point[44] and the higher vapor pressure of Mo species [5], $MoSe_2$ nucleated first, followed by epitaxial $WSe_2$ growth from the $MoSe_2$ edges. At a higher Mo concentration (3:1), a bilayer $MoSe_2$ formed and thereby promoted the growth of vertical HS II. In contrast, a lower Mo concentration (1:3) favored the growth of ML $MoSe_2$ and lateral $WSe_2$ stitching to form HS I. Individual $MoSe_2$ growth at high and low concentrations of $Na_2MoO_4$, representing the 3 and 1 values in the precursor ratios, confirms this mechanism yielding mainly bilayer or ML $MoSe_2$, respectively (*Figure S4*). A heating rate of 15 °C min$^{-1}$ was essential for pure phase of $MoSe_2$-$WSe_2$ LHs formation as faster heating rates (40 °C min$^{-1}$) led to mixed-phase LHs (*Figure S5*), which is consistent with the fast kinetics of TMD MLs at our experimental conditions[44, 45]. Although the growth temperature did not affect preferential growth of HS I or HS II, at higher temperatures lateral sizes of HS I are increased (*Figures S6-S7*). In addition, increasing the W precursor concentration expanded WSe$_2$ domains by approximately two to threefold (*Figure S7*), with the entire lateral sizes of HS I ranging from ~10-200 µm (*Figures S7-S8*) and of top LH of HS II ranging from ~20-60 µm (*Figures S6, S9*). Thus, using liquid precursors, both HS I and HS II can be grown uniformly across the substrate (*Figures S8a and S9a*); further technical details of the HS growth are presented in the Experimental Section.

**2.2 Spectroscopic and microscopic characterization of HS I and HS II down to the nanoscale**

To study the properties of the formed HS I and HS II, we applied several complementary microscopic and spectroscopic techniques at room temperature (RT). We first conducted Raman spectroscopy on the inner and outer regions of lateral HS I to identify the MLs and their structural quality, *Figure 2a*. The inset of *Figure 2a* shows an optical microscopy image indicating the measurement locations. The characteristic Raman peaks at 240 cm$^{-1}$ (A′$_1$ mode) and 286 cm$^{-1}$ (E′ mode) confirm the presence of MoSe$_2$ in the inner region. Peaks at 249 cm$^{-1}$ (E′/A′$_1$ modes) and 260 cm$^{-1}$ (2LA mode) correspond to the WSe$_2$ in the outer region. The narrow full width at half maximum (FWHM) values of the A′$_1$ mode of MoSe$_2$ (1.7 cm$^{-1}$) and the E′/A′$_1$ modes of WSe$_2$ (3.7 cm$^{-1}$) indicates high crystalline quality of the HSs grown from liquid precursors, which is comparable to those grown from solid precursors of transition metals[5]. To assess the uniformity of these lateral HSs, Raman intensity maps were recorded for the A′$_1$ mode of MoSe$_2$ and the E′/A′$_1$ mode of WSe$_2$ across the entire MoSe$_2$-WSe$_2$ HS I (*Figures 2b and 2c*). The A′$_1$ peak intensity map of MoSe$_2$ (*Figure 2b*) reveals a uniform signal within the inner region of the HS, while the E′/A′$_1$ peak intensity map of WSe$_2$ (*Figure 2c*) shows localization in the outer region only. These results confirm the well-defined formation of lateral HS I. Further, AFM studies show (*Figure S10*) a thickness of ~0.7 nm, indicating a ML structure.

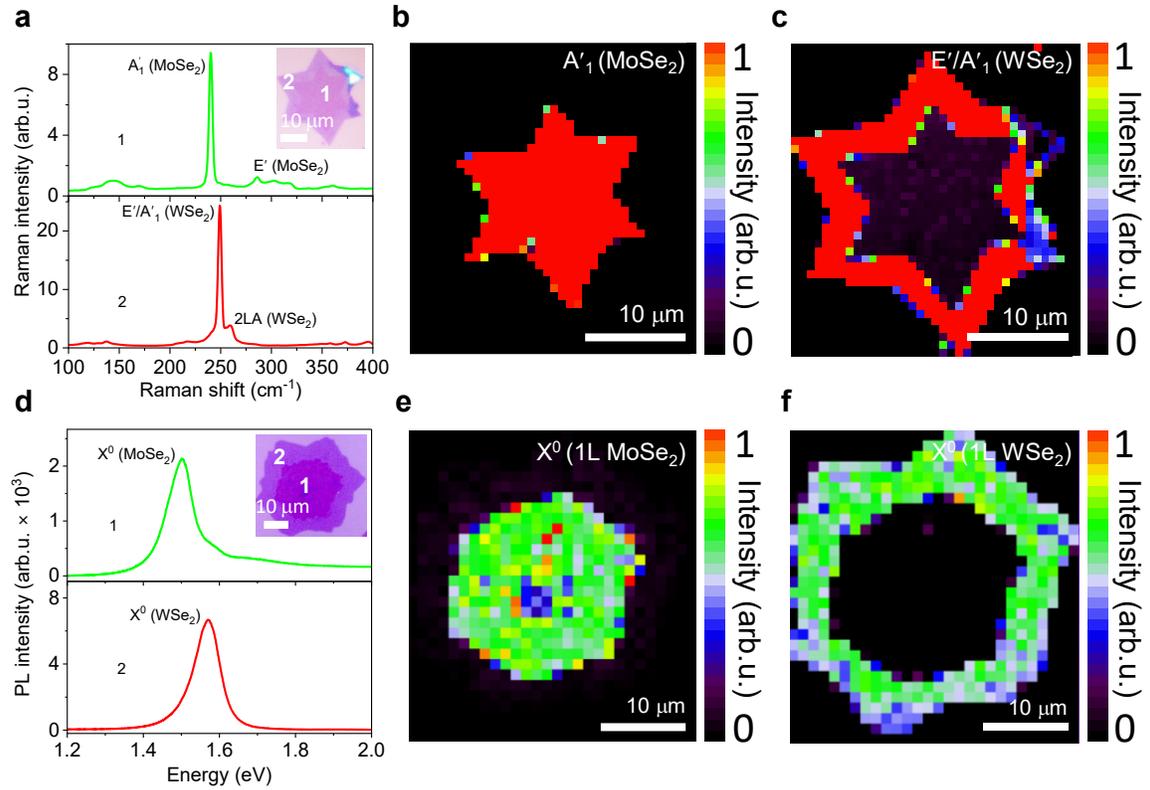

**Figure 2:** Raman spectroscopy and PL characterization of HS I at RT. (a) Raman spectra acquired from region 1 (MoSe$_2$) and region 2 (WSe$_2$) with the inset displaying the optical microscopy image indicating the measurement positions. (b) Raman intensity map of A′$_1$ mode (MoSe$_2$) recorded over the area shown in the inset of (a), showing uniform crystallinity in the inner region. (c) Raman intensity map of the E′/A′$_1$ mode (WSe$_2$) showing uniform crystallinity in the outer region. (d) PL spectra of inner MoSe$_2$ (bottom panel) and outer WSe$_2$ (top panel). The inset shows the optical microscopy image indicating the corresponding measurement positions. (e, f) PL intensity map of MoSe$_2$ and WSe$_2$, collected over the area shown in the inset of Figure (d).

To evaluate the optical properties of the lateral HS I, we performed RT photoluminescence (PL) measurements using a 532 nm laser. The PL spectra (*Figure 2d*) show a ML MoSe$_2$ emission peak at 1.50 eV (neutral exciton, $X^0$) and a ML WSe$_2$ emission peak at 1.57 eV ($X^0$). Both emission energies are significantly lower than typically reported values of 1.57 eV (MoSe$_2$) and 1.65 eV (WSe$_2$)[3], suggesting that the heterostructures are under tensile strain, which was likely introduced during the CVD process due to the different substrate/TMD thermal expansion coefficients.[46] Despite this strain effect, the strong PL intensity indicates that HS I remains a direct bandgap and of high optical quality. The PL intensity of the WSe$_2$ region is approximately 3.5 times stronger than that of MoSe$_2$, consistent with a previous report on lateral MoSe$_2$-WSe$_2$ HS grown from solid precursors (mixtures of MoO$_3$ and WO$_3$ powders)[47]. *Figures 2e, f* present spatial PL maps of lateral MoSe$_2$-WSe$_2$ HS I. The PL mapping at 1.50 eV (*Figure 2e*) clearly highlights the inner MoSe$_2$ region, while the 1.57 eV map (*Figure 2f*) identifies the outer WSe$_2$ region.

To elucidate the nanoscale interface structure of HS I, we performed HAADF-STEM, where images are sensitive to atomic-number (Z) contrast, in combination with energy-dispersive X-ray spectroscopy (EDX) elemental mapping (*Figures 3a-c and S12*). As expected, the EDX elemental maps (*Figure 3b* and *Figure S12*) show a uniform distribution of Se across the entire HS I, while Mo is localized in the centre and W in the outer regions. The contrast differences in the overview HAADF-STEM image (*Figure 3a*, darker contrast for lighter Mo atoms and brighter contrast for heavier W atoms), clearly supports the EDX results and reveals additional details – an intermediate Mo-W-Se region of approximately 120 nm in width between MoSe$_2$ and WSe$_2$, where both transition metals are present, indicating alloy formation. High resolution HAADF-STEM further shows that the interfaces between MoSe$_2$/Mo-W-Se (*Figure 3c*) and Mo-W-Se/WSe$_2$ (*Figure S13*) are confined to only ≈ 5 nm. Selected-area electron diffraction (SAED) patterns acquired from these three distinct regions (*Figure 3d-f*) show sharp six-fold symmetry, and with nearly identical {10-10} d-spacings, demonstrating a single crystalline in-plane heterojunction in HS I.

Next, we characterize the properties of HS II samples. As introduced in Section 2.1, HS II is a bilayer configuration with a bottom MoSe$_2$ ML and a lateral MoSe$_2$-WSe$_2$ top layer. Raman spectroscopy (see *Figure 4a*) at three different regions (marked with 1, 2, and 3 in the inset of *Figure 4a* see also *Figure 1d*) identifies the layers of such HS II. In region 1 (bottom layer), the Raman modes at 240 (A′$_1$) cm$^{-1}$ and 286 (E′) cm$^{-1}$ confirm the MoSe$_2$ ML, consistent with the peak positions observed in ML MoSe$_2$ of HS I. In region 2 (heterobilayer region), the Raman modes at 241 (A′$_1$ of MoSe$_2$) cm$^{-1}$, 249 (E′/A′$_1$ of WSe$_2$) cm$^{-1}$, and 259 (2LA of WSe$_2$) cm$^{-1}$ confirm the vertical HS of MoSe$_2$/WSe$_2$. In region 3 (homobilayer region), the characteristic Raman modes at 241 (A′$_1$) cm$^{-1}$ and 286 (E′) cm$^{-1}$ confirm the MoSe$_2$. Noticeably, the A′$_1$ mode of MoSe$_2$ (compared between regions 1 and 3) shows a discernible blueshift, as expected in bilayer MoSe$_2$ because of weak van der Waals interactions.[48] Raman intensity maps of the A′$_1$ mode of MoSe$_2$ and E′/A′$_1$ mode of the WSe$_2$ on the entire lateral/vertical MoSe$_2$-WSe$_2$ HS II are depicted in *Figures 4b* and *4c*, respectively. The intensity of the A′$_1$ peak of MoSe$_2$ (*Figure 4b*) is primarily detected in regions 1 and 3 of the HS, with a lower contrast in region 2, as expected. The intensity map of the E′/A′$_1$ peak of WSe$_2$ (*Figure 4c*) shows that WSe$_2$ is only present in region 2, validating the vertical MoSe$_2$/WSe$_2$ HS formation. The AFM height profile across the bottom MoSe$_2$ ML and top lateral MoSe$_2$-WSe$_2$ HS (*Figure S14*) shows an estimated height of ∼ 0.7 nm, further confirming the ML of lateral MoSe$_2$-WSe$_2$ on top of the MoSe$_2$ ML.

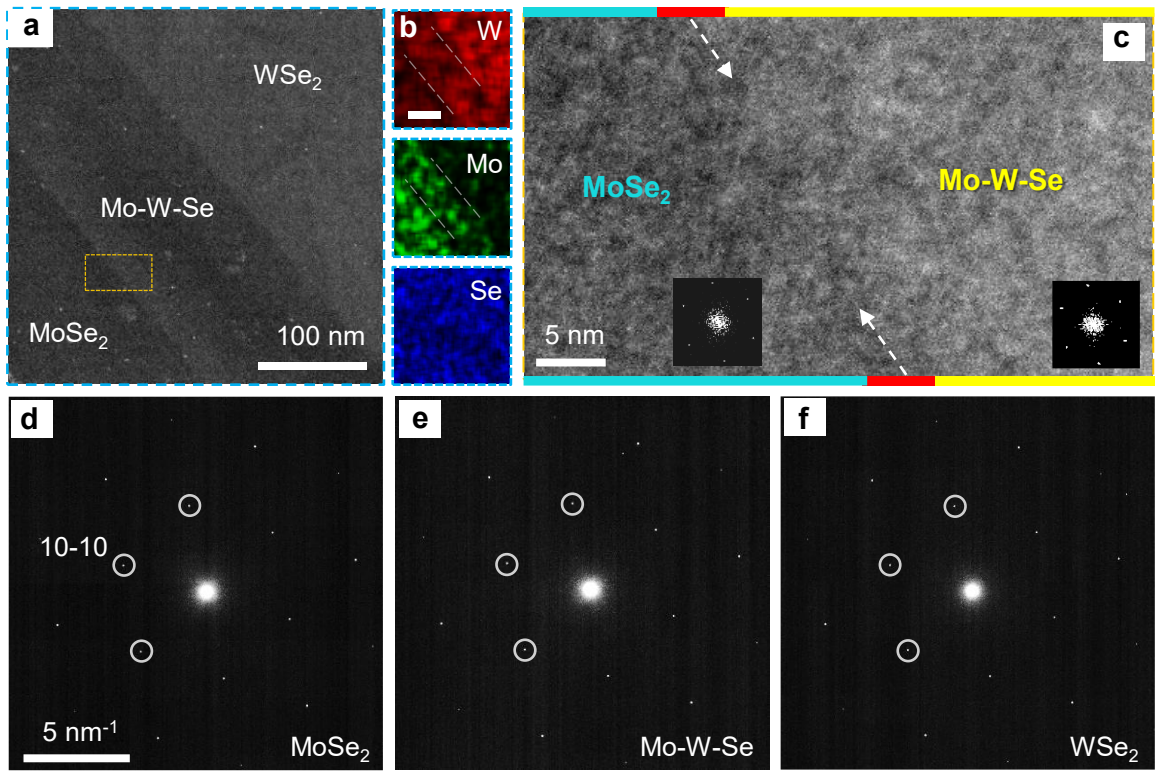

**Figure 3:** Nanoscale characterization of HS I interface. (a) 200 kV HAADF-STEM image of the heterostructure interface, revealing an ~120 nm Mo-W-Se region bridging the $MoSe_2$ and $WSe_2$ domains. (b) Elemental maps obtained from (a) of W, Mo, and Se highlight the compositional transition across the intermixed, alloyed region. (c) High-magnification 200 kV HAADF-STEM image of the $MoSe_2$ and Mo-W-Se boundary shows a narrow ~5 nm interface (red lines), with $MoSe_2$ and Mo-W-Se regions indicated by cyan and yellow lines, respectively. A white dashed arrow is included as a guide to the eye. The inset displays the corresponding fast Fourier transforms (FFTs) obtained from the respective regions. (d-f) SAED patterns obtained at 200 kV with the smallest SAED aperture of 10 μm resulting in an about 100 nm wide area from the $MoSe_2$, Mo-W-Se, and $WSe_2$ regions, respectively. The {10-10} plane d-spacings are 0.2818 nm ($MoSe_2$), 0.2819 nm (Mo-W-Se), 0.2821 nm ($WSe_2$) (see the $WSe_2$ HAADF image in *S13*).

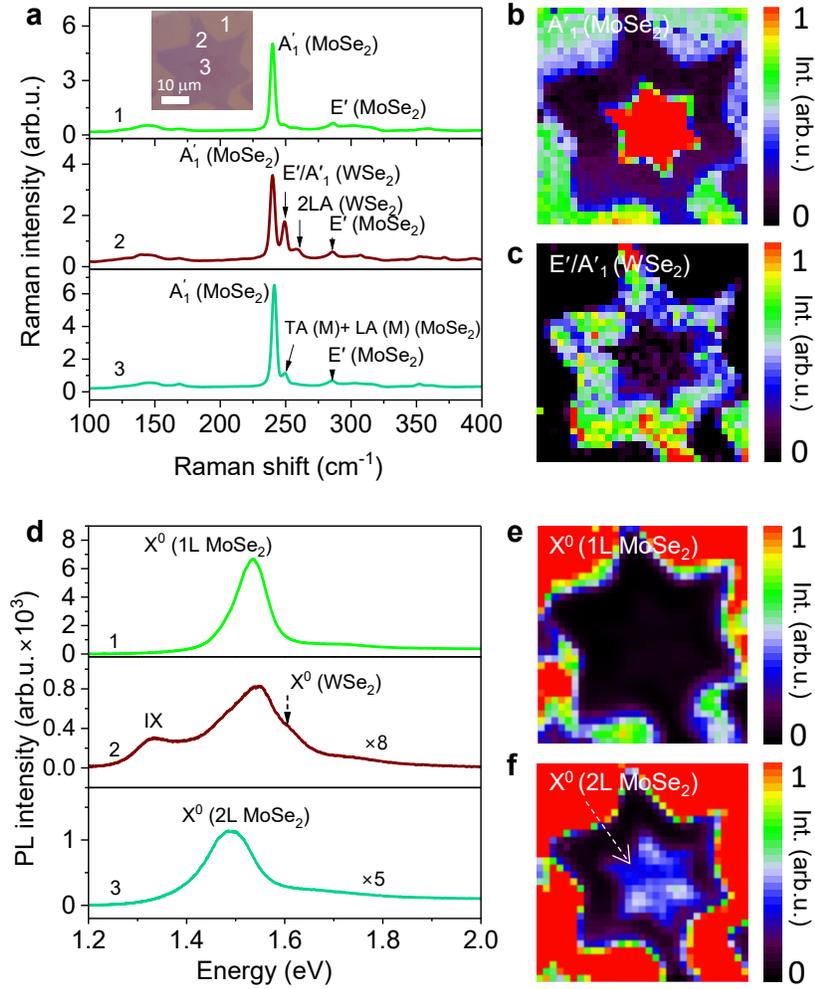

**Figure 4:** Raman spectroscopy and PL characterization of HS II at RT. (a) Raman point spectra at regions 1 (1L MoSe$_2$), 2 (vertical MoSe$_2$-WSe$_2$ bilayer), and 3 (2L MoSe$_2$) as shown in the optical microscopy image of hybrid lateral-vertical MoSe$_2$-WSe$_2$ HS II in the inset. (b) and (c) show the Raman intensity spatial maps of A$'_1$ of MoSe$_2$ and E$'$/A$'_1$ of WSe$_2$, respectively. (d) PL spectra at the three phases, which were collected from three different regions of the inset of Figure (a). (e) and (f) show the corresponding PL intensity map of 1L MoSe$_2$ (1.54 eV) and 2L MoSe$_2$ (1.49 eV) in the PL spectra in Figure (d).

Next, we analyse the PL spectra of HS II. As seen from *Figure 4d*, the PL intensity at 1.54 eV ($X^0$) from bottom ML MoSe$_2$ (region 1) in comparison to that at 1.49 eV ($X^0$) from bilayer MoSe$_2$ (region 3) is ~ 6 times higher as expected for a direct and non-direct band gap semiconductors[49], respectively. From the value of the $X^0$ emission energy for the ML MoSe$_2$ some substrate induced tensile strain is expected[50], which is lifted if HS II is encapsulated in hBN (see Section 2.3). Most prominent that in region 2 the PL spectrum reveals also the IE emission at ~1.33 eV (IX), see middle spectrum in *Figure 4d* and *Figure S15*, due to type-II band alignment between MoSe$_2$ and WSe$_2$.[3] Additionally, *Figures 4e* and *4f* represent the PL mapping of the HS II at 1.54 eV and at 1.49 eV, which clearly reflect regions 1, 2, and 3 of HS II heterostructure.

To characterize the nanoscale structure of HS II, we performed HAADF-STEM imaging combined with EDX analysis (*Figure 5* and *Figure S16*). The EDX data from region 2 reveal nearly equal atomic fractions of Mo and W, consistent with a vertical heterobilayer of MoSe$_2$ and WSe$_2$, whereas regions 1 and 3 are dominated by Mo signal. Examination of the MoSe$_2$-WSe$_2$ lateral interface shows an interface width of only ≈ 5 nm (*Figure 5a-c*). We attribute this sharp interface to the epitaxial growth of the top lateral HS on the bottom MoSe$_2$ monolayer. This interpretation is supported by the continuous atomic rows with no visible Moiré pattern in the HRTEM image (*Figure 5d and S17a*) and by the SAED patterns showing a single set of sharp circular Bragg spots (*Figure 5e, S18*), both of which are a direct evidence of epitaxial alignment. In contrast, non-epitaxial growth of the 2$^{nd}$ layer, or a twisted bi-layer, would typically produce Moiré-fringes in HRTEM images and split diffraction spots in the SAED.

A small lattice mismatch between MoSe$_2$ and WSe$_2$ facilitates reconstruction phenomena in the vertical stacks of the constituent MLs, giving rise to rich excitonic features.[51] We employed scanning electron microscopy (SEM) to investigate the mesoscopic reconstruction of MoSe$_2$/WSe$_2$ heterobilayer region in HS II on a SiO$_2$/Si substrate. The SEM image (*Figure 5f*) reveals a pattern of one-dimensional (1D) stripes, characteristic of such reconstruction.[51] To further determine the stacking order, we combined HAADF, annular dark field (ADF) imaging, and localized SAED measurements (*Figure 5 (g-j)*). Typically, HAADF contrast is dominated by Z-contrast[52], whereas ADF contrast [53], and the relative diffraction spots intensities in SAED provide insight into stacking order[54, 55]. In both HAADF and ADF images (see *Figures 5g,h*), we observe 1D stripes consistent with the reconstructed pattern seen in SEM (*Figure 5f*). In addition, the ADF image reveals brighter and darker domains (*Figure 5h*) that are not clearly resolved in the HAADF image (*Figure 5g*), indicating that the top WSe$_2$ ML forms domains of different stacking orders on the bottom MoSe$_2$ ML. We label these domains α (brighter) and β (darker) and acquired the SAED patterns from each domain. The SAED pattern from domain α (*Figure 5i*) shows comparable intensities of the first- and second-order diffraction spots (inset of *Figure 5i*), consistent with AB (H-like) stacking.[54, 55] In contrast, the SAED pattern from domain β (*Figure 5j*) exhibits markedly weaker first-order diffraction intensities relative to the second-order spots (inset of *Figure 5j*), indicatiing AA (R-like) stacking.[54, 55] High-resolution TEM imaging (*Figure S19*) further supports the coexistence of AB (H-like) and AA (R-like) stacking, demonstrating that the heterobilayer in HS II contains a hybrid of both stacking domains. Thus, the heterobilayer exhibits mesoscopic reconstruction and mixed stacking domains.

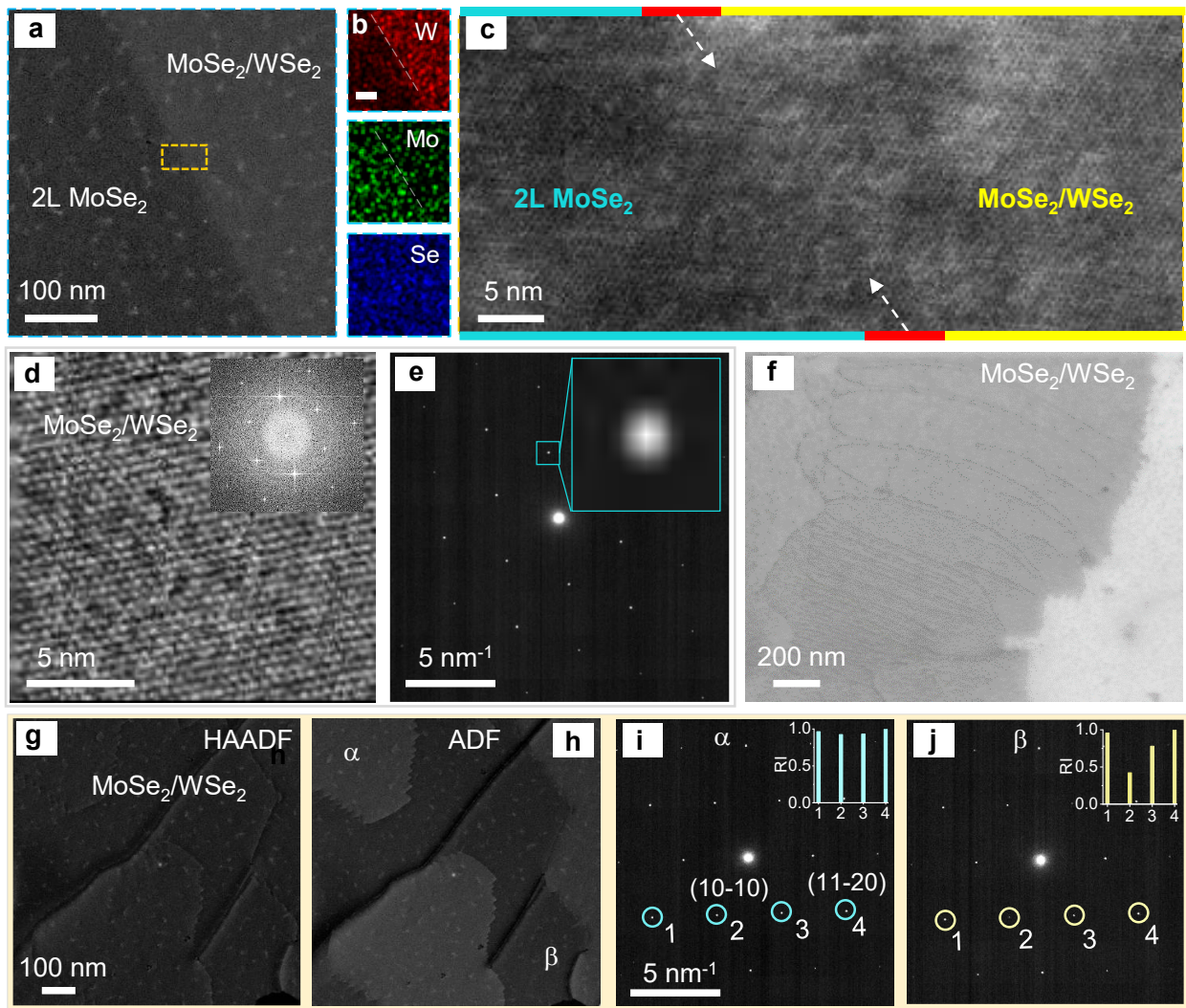

**Figure 5:** Nanoscale characterization of HS II structure. (a) 200 kV HAADF-STEM image of the interface of the LH region. (b) EDX elemental maps of W, Mo, and Se of the image in (a). (c) 200 kV high-resolution HAADF-STEM image showing the ~5 nm interface between the LH MoSe$_2$-WSe$_2$ regions. (d) 200 kV HRTEM image of the vertical MoSe$_2$/WSe$_2$ heterostructure region. The inset represents the fast Fourier transform (FFT). (e) The SAED pattern acquired at 200 kV exhibits six-fold symmetry with single circular diffraction spots, which indicates epitaxial growth of the top LH. The inset shows a magnified view of a single diffraction spot. (f) SEM micrograph of the MoSe$_2$/WSe$_2$ region of HS II on 300 nm SiO$_2$/Si substrates reveals the reconstruction patterns. (g, h) Comparison of 200 kV HAADF- and ADF-STEM images of the MoSe$_2$/WSe$_2$ region, where domains of brighter and darker contrast are clearly resolved in the ADF image, labelled as α and β, respectively. (i, j) Corresponding 200 kV SAED patterns at domains α and β with insets showing intensities of first (positions 2,3)- and second (positions 1,4)-order diffraction spots, respectively, indicating the H (α) and R (β)- like stacking domains.

## 2.3 Strong interlayer exciton emission at MoSe₂/WSe₂ heterobilayer in HS II across 4 K - 300 K range

Previous work on CVD grown samples has shown the benefit of hBN encapsulation.[56-58] As the samples are lifted off their growth substrates, the layers are essentially unstrained.[59] This is evidenced by the room-temperature PL spectral map of the hBN-encapsulated HS II (see *Figure S20 and S21*), which shows uniform, strain-free emission energies across the heterostructure. To investigate the heterostructure optically, we have performed PL measurements. In *Figure 6a*, we show the PL emission from the MoSe₂/WSe₂ heterobilayer region at T = 4 K. Here a distinct emission corresponding to interlayer excitonic species is observed. These peaks labelled $IX_T^-$ (1.395 eV) and $IX_T$ (1.402 eV) are attributed to the charged and neutral spin-triplet interlayer excitons.[40, 60] In this region of the sample, no emission from intralayer species is discernible. This dominance of the interlayer emission indicates very efficient transfer of electrons from WSe₂ and holes from MoSe₂, leading to the formation of IEs (as represented in the inset of *Figure 6a*). Such a strong signature of the IE is a promising characteristic of liquid precursor-grown heterobilayers. *Figure 6b* shows the PL emission from 1L-MoSe₂ region at T = 4 K where the contributions from trion ($X_{MoSe_2}^-$) and neutral ($X_{MoSe_2}^0$) excitons.[61, 62] As a control, we compared the PL properties of our CVD grown MoSe₂/WSe₂ heterostructure in HS II with those of an exfoliated MoSe₂/WSe₂ heterostructure (*Figure S22*). The CVD-grown sample exhibits significantly narrower interlayer exciton peaks, with FWHM values reduced by ~ 37 % for the $IX_T^-$ ($IX_T^-(CVD)$ = 6.8 ± 0.13 meV vs. $IX_T^-(Exf.)$ = 10.78 ± 0.57 meV) and ~ 67 % for the $IX_T$ ($IX_T(CVD)$ = 4.27 ± 0.06 meV vs. $IX_T(Exf.)$ = 13.03 ± 0.21 meV), compared to the exfoliated sample. These results highlight the superior optical quality and atomically cleaner vertical interface of the CVD-grown HS. In addition, ~ 54 % linewidth reduction in the 1L-MoSe$_2$ intralayer exciton (from 5.67 ± 0.12 meV to 2.6 ± 0.02 meV) indicates excellent optical quality even in the ML MoSe$_2$ region 1 of HS II.

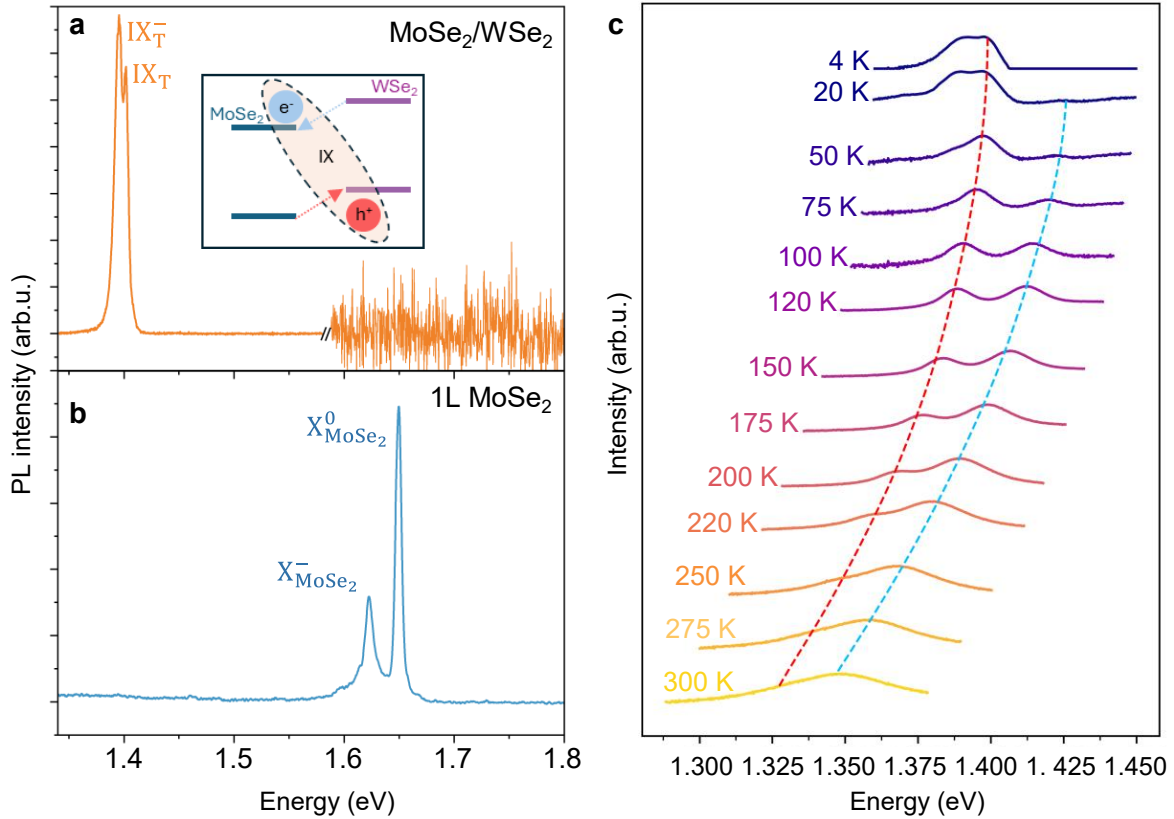

**Figure 6:** PL spectroscopy of interlayer excitons in hBN encapsulated HS II: Emission at T = 4 K from (a) the MoSe$_2$/WSe$_2$ heterobilayer showing the neutral ($IX_T$) and charged ($IX_T^-$) spin-triplet interlayer excitons and (b) the 1L-MoSe$_2$ with contributions from the trion ($X_{MoSe_2}^-$) and neutral ($X_{MoSe_2}^0$) exciton species. (c) Temperature dependent PL from the heterobilayer region with Varshni fits. $IX_T$ shows a discernible emission from 4 K to 300 K. $IX_T^-$ vanishes above 20 K. Above 20 K additional peaks appear in the interlayer emission spectral region owing to the spin-singlet state $IX_S$. At temperatures below 75 K the PL emission is dominated by $IX_T$, however at higher temperatures, the singlet state, $IX_S$ dominates the emission.

Most studies on IEs focus on spectroscopy at cryogenic temperatures. But in *Figure 4d* a spectral feature at the IE energy seems to be discernible at RT. This motivated our temperature dependent measurements to follow the evolution of the PL emission from T = 4 K to 300 K. The results are summarized in *Figure 6c*. It can be seen that we are indeed able to trace the IE emission at all temperatures between 4 K and 300 K. For temperatures between 4 K and 75 K, the contribution to the PL is mainly from emissions IX$_T^-$ and IX$_T$. At temperatures above 20 K, a higher energy contribution corresponding to the spin-singlet interlayer exciton (IX$_s$) is detected.[40] From around 50 K onward, the emission of IX$_T^-$ is no longer detectable, possibly indicating weak localization.[63] Following the interlayer emission through increasing temperatures, it is observed that the singlet state IX$_s$ gains intensity and becomes the dominant peak at 300 K. This has certain parallels to the temperature evolution of the intralayer exciton PL in monolayer WSe$_2$, where the spin-allowed bright exciton state becomes more intense with temperature[64]. The singlet state IX$_s$ emission also corresponds to the RT transition energy observed in *Figure 4d*. Globally, the IE emission dominates the PL in region 2 of HS II for all temperatures investigated. Furthermore, as can be seen in *Figure 6c*, the individual emission energies are fitted with Varshni functions.[65, 66] The fits are indicated as dashed lines in the plot. All the peak shifts are in good agreement with the Varshni function. Thus showing that we can track the IE emission over the entire temperature range. This excludes defects to be at the origin of the low temperature PL. The temperature evolution of the IE emission energy depends on the valence band shift in WSe$_2$ and the conduction band shift in MoSe$_2$. The Varshni parameters for intralayer MoSe$_2$, WSe$_2$, and IEs are compared in Table S2 (Supplementary Information). The PL characteristics reported in this section correspond to 2H stacking for the sample investigated, as the attribution of exciton lines in excellent agreement with the literature[40, 60]

**Conclusions**

In summary, we demonstrate a scalable bottom-up CVD methodology, based on liquid transition metal salts precursors, for the controlled growth of two high-quality MoSe$_2$-WSe$_2$ heterostructures: purely lateral (HS I) and hybrid lateral/vertical (HS II). By simply adjusting precursor concentration, we selectively direct the preferential growth of HS I and HS II structures. We found that the vertical MoSe$_2$/WSe$_2$ heterobilayers in HS II exhibit strong interlayer excitonic emission that remains visible at RT and becomes the sole contributor at 4 K – a promising feature that indicates exceptionally clean and atomically sharp interfaces of these samples. We envision that our simple CVD growth process of HS I and HS II from liquid precursors will enable the rapid synthesis of these complex heterostructures for excitonic devices, which can readily offer a scalable alternative to manually assembled heterostructures and can be integrated into nanoelectronic and photonic devices.

**Experimental**

*Sample preparation*

*Materials:* Sodium molybdate ($Na_2MoO_4$ powder, 98.0%, Sigma Aldrich), sodium tungsten oxide dihydrate ($Na_2WO_4$, $2H_2O$ powder, 99.0%, Alfa Aesar), and selenium granules (99.9%, Sigma Aldrich) were used without further purification.

*Aqueous sodium molybdate and sodium tungstenate solution preparation:* For the molybdenum (Mo) and Tungsten (W) source, the aqueous solution of sodium molybdate and sodium tungstenate was prepared in ultrapure water (0.056 µS/cm, MembraPure Aquinity2 E35).

*Precursor substrates preparation:* ~ 1.5 cm × 1.5 cm Si substrate with a 300 nm thick $SiO_2$ layer (Sil'tronix, root mean square (rms) roughness <0.2 nm) was used as growth substrate. To improve the substrate's hydrophilicity and increase the wettability of Mo/W precursor during spin coating, the substrates were exposed to $O_2$ (10 standard cubic centimeters per minute (sccm)) plasma for 280 seconds (Diener Zepto). Further, the aqueous solution of sodium molybdate and sodium tungstenate in a 1:3, 1:10, and 3:1 ratio was spin-coated at 2000 rpm on plasma-cleaned $SiO_2$/Si substrate, followed by heating at 70°C for 5 minutes.

*Controlled synthesis of $MoSe_2$-$WSe_2$ HS I and HS II:* A two-zone furnace equipped with a quartz tube of 55 mm diameter (Carbolite Gero) was used for the growth in a similar manner as reported in. Ref. [43]. The growth substrate was heated slowly (15 °C $min^{-1}$) to different growth temperatures in the range of 770-900 °C and maintained for 15 minutes to grow the HS I (at 1:3 and 1:10) and HS II (at 3:1).

*Transfer of Heterostructure on TEM grid:* The *HS II* were transferred onto the lacey carbon-coated Cu TEM grid via the wet transfer method.[67] In brief, a layer of poly(methyl methacrylate) (PMMA) was spin-coated onto the as-grown HS. Thereafter, the sample was floated on top of potassium hydroxide aqueous solution (2 M) to delaminate the PMMA/HS film from the $SiO_2$/Si substrate and subsequently washed with ultrapure water to remove residual KOH. Afterwards, the PMMA/HS film was placed onto the TEM grid and baked at 90 °C for 10 min to promote adhesion between the HS and lacey carbon on Cu. Next, the PMMA/HS/grid samples were introduced into a $CO_2$ critical point dryer (CPD, Tousimis)(a similar description is available in [68]) to dissolve PMMA and to achieve a contamination-free clean surface of the HS before introducing it into the electron microscope. For the CPD, the PMMA/HS/grid samples were kept in acetone for 45 min in the process chamber. The acetone was replaced by liquid $CO_2$ at low temperature (~ -2 °C), maintaining a pressure of 5 MPa. To get the supercritical $CO_2$ condition, the temperature was raised to 38 °C at a pressure of 10 MPa for 10 min. The HS/grid samples were taken out of the CPD chamber once the temperature and pressure reached ambient conditions. The transfer of HSs onto an Au Quantifoil TEM grid was performed without PMMA, using isopropanol and KOH as transfer agents.

*Characterization*

*Optical Microscopy (OM):* Axio Imager Z1.m microscope (Zeiss) equipped with a thermoelectrically cooled 3-megapixel CCD camera (Axiocam 503 color) was used to capture the OM images in bright-field operation.

*Atomic Force Microscopy (AFM):* AFM measurements were performed with a Ntegra system (NT-MDT) in tapping mode at ambient conditions using n-doped silicon cantilevers (NSG01, NT-MDT) with a tip radius of 10 nm with resonant frequencies of 167 kHz.

*Raman and Photoluminescence (PL) Spectroscopy:* Raman spectra were acquired using a Bruker Senterra spectrometer operated in backscattering mode. A 532 nm laser was used to excite the sample under a 100x objective, and a thermoelectrically cooled CCD detector collected the Raman signals through 1800 l/mm grating. The spectral resolution of the system is 2-3 cm$^{-1}$. The Si peak at 521 cm$^{-1}$ was used for calibration of the instrument. For PL spectra, the same 532 nm (2.33 eV) excitation was used with a grating of 1800 l/mm. Raman spectroscopy measurements were performed using a 2 mW excitation power and an integration time of 2 seconds. Raman and PL mapping were performed using a Renishaw inVia Raman microscope, operated in backscattering mode with a 532 nm excitation laser (Nd:YAG laser). A 100× objective lens with NA 0.85 was used for all measurements, and the system was equipped with an 1800 l/mm grating. The sample stage was motorized to enable automated mapping, and data were collected using a CCD detector. The acquired mapping data were analyzed using Renishaw's Wire 5.6 software.

*Scanning electron microscopy (SEM):* SEM images were acquired using a Zeiss Sigma VP scanning electron microscope operated at an accelerating voltage of 2 kV, using the in-lens detector.

*Transmission electron microscopy including (HRTEM, STEM, and EDX mapping):* Prior to TEM analysis, the samples were removed from the substrate and transferred onto the TEM grids coated with lacey carbon films. High-resolution TEM (HRTEM), high-angle annular dark-field scanning TEM (HAADF-STEM) imaging and energy dispersive X-ray (EDX) mapping were acquired on a Thermo Fisher Talos 200X equipped with a SuperX four quadrant EDX detector. The microscope was either operated at 80 kV or 200 kV with a probe-current of 100 to 200 pA in STEM mode and about 5 nA in TEM mode. 80 kV accelerating voltage was used initially because of reduced knock-on damage and higher inelastic scattering improving the EDX sensitivity. However, at 200 kV the sample contamination in STEM mode is reduced and the spatial resolution in TEM and STEM mode is higher. The EDX energy dispersion was set to 5 eV per channel over an energy range of 0 to 20 keV. For EDX analysis, multiple drift-corrected single-frame scans (512 x 512 pixels) were taken and summed up. Typical mapping times were about 10 min; Owning to the low signal-to-noise ratio from monolayer regions, faint signals outside the flake originate from background noise only.

*Encapsulation of as-grown samples:* To provide the samples with a clean and strain-free dielectric environment, they are encapsulated in layers of hexagonal boron nitride (hBN). A hot pick-up technique[38] using polydimethylsiloxane (PDMS) and poly(bisphenol-A) carbonate (PC) is employed for the encapsulation. Following the removal of the PC film, the encapsulated sample was annealed at 110°C for 20 minutes.

*Optical Spectroscopy of hBN encapsulated samples:* For the optical spectroscopy carried out between T = 4K and 300 K the samples were placed in a attoDRY800 ultra-low vibration, closed-cycle cryostat system and a home-built microscopy setup was used. The spectra were acquired using a Czerney-Turner spectrograph equipped with a CCD Camera. For PL at T = 4 K, the excitation laser power was 20 µW, with an acquisition time of 20 ms. For the temperature-dependent PL measurements, the laser power was kept constant at 20 µW, while the acquisition time varied between 20 ms and 5000 ms; all data were normalized to /s. A 533 nm excitation laser was used, with a laser spot size of approximately 580 nm. We used low-temperature piezo-positioners (attocube systems, ANPx101and ANPz102) to position the sample and perform line scans. To achieve a diffraction-limited excitation spot low-temperature apochromatic objective was used. PL measurements were performed in backscattering geometry.

## ASSOCIATED CONTENT

### Supporting Information

It includes additional characterization of HS I and HS II heterostructures using OM, AFM, PL, and TEM, along with growth temperature profiles. Supplementary tables provide precursor ratios, molar concentrations, Varshni parameters for ML MoSe$_2$, and IX$_T$ and IX$_S$ features from HS II and exfoliated ML WSe$_2$.

### Author Contributions

The manuscript was written through contributions of all authors. All authors have given approval to the final version of the manuscript.

### Notes

The authors declare no competing financial interest.

## ACKNOWLEDGMENTS

T.H., A.P., C.N and A.T. acknowledge the financial support of this work via the European Union Graphene Flagship project 2D Materials of Future 2DSPIN-TECH (No. 101135853), by the Deutsche Forschungsgemeinschaft (DFG, German Research Foundation) - CRC/SFB 1375 NOA "Nonlinear Optics down to Atomic scales" (Project B2, number 398816777), by BMBF Project SINNER Grant No. 16KIS1792, SPP2244 "2DMP" (Project TU149/21-1, 535253440), and DFG individual grant TU149/16-1 (464283495). U.K. and J.B. acknowledge the financial support of the DFG via the grant KA1295/45-1 (464283495). S.Sh and B.U acknowledge financial support from DFG UR 312/2-1 Projektnummer 535253440. T.H. acknowledges Gabriele Es-Samlaoui for assistance with TEM sample preparation. We thank Stephanie Höppener and Ulrich S. Schubert for enabling Raman spectroscopy and scanning electron microscopy studies at the Jena Center for Soft Matter (JCSM).

# Supporting Information

# CVD Grown Hybrid MoSe$_2$-WSe$_2$ Lateral/Vertical Heterostructures with Strong Interlayer Exciton Emission

*Md Tarik Hossain[1]\*, Sai Shradha[2], Axel Printschler[1], Julian Picker[1], Luc F. Oswald[2], Julian Führer[2], Nicole Engel[2], Honey Jayeshkumar Shah[1], Christof Neumann[1], Daria I. Markina[2], Moritz Quincke[3], Johannes Biskupek[3], Kenji Watanabe[4], Takashi Taniguchi[5], Ute Kaiser[3], Bernhard Urbaszek[2], Andrey Turchanin[1,6]\**

[1]Institute of Physical Chemistry, Friedrich Schiller University Jena,
Lessingstr. 10, 07743 Jena, Germany

[2]Institute of Condensed Matter Physics, Technical University Darmstadt,
Hochschulstraße 6-8, 64289 Darmstadt, Germany

[3]Central Facility of Electron Microscopy, Electron Microscopy Group of Material Science, Ulm University, 89081 Ulm, Germany

[4]Research Center for Electronic and Optical Materials,
National Institute for Materials Science, 1-1 Namiki, Tsukuba 305-0044, Japan

[5]Research Center for Materials Nanoarchitectonics,
National Institute for Materials Science, 1-1 Namiki, Tsukuba 305-0044, Japan

[6]Abbe Center of Photonics, Albert-Einstein-Straße 6, 07745 Jena, Germany

**Corresponding Authors**:

Md Tarik Hossain (tarik.hossain@uni-jena.de),

Andrey Turchanin (andrey.turchanin@uni-jena.de)

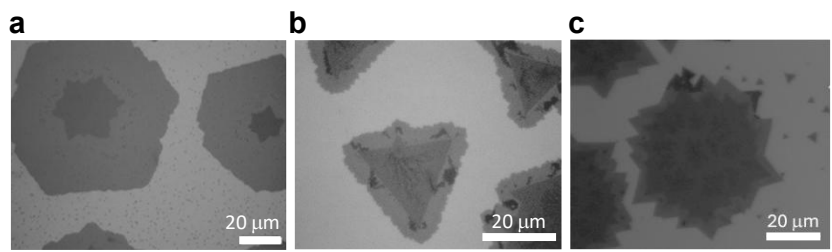

**Figure S1:** Optical microscopy images of differently shaped lateral $MoSe_2$-$WSe_2$ HS I. (a) hexagonal (b) triangle (c) jagged edged.

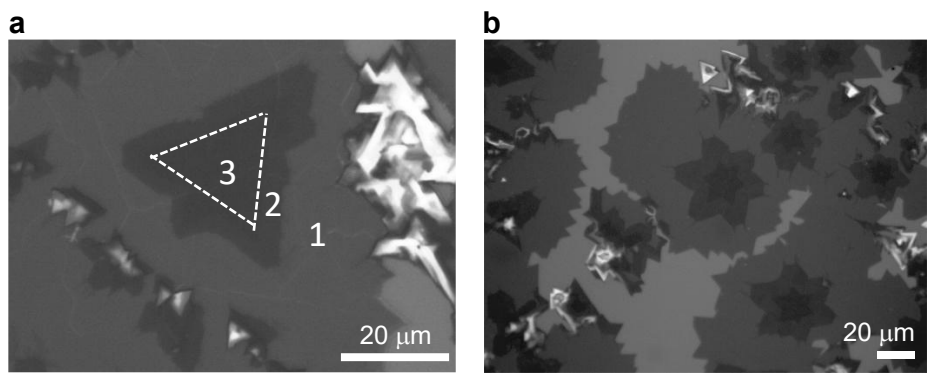

**Figure S2:** Optical microscopy images of a differently shaped $MoSe_2$-$WSe_2$ HS II (lateral/vertical heterostructure). (a) triangle, (b) hexagon on the bottom jagged-edged $MoSe_2$ monolayer.

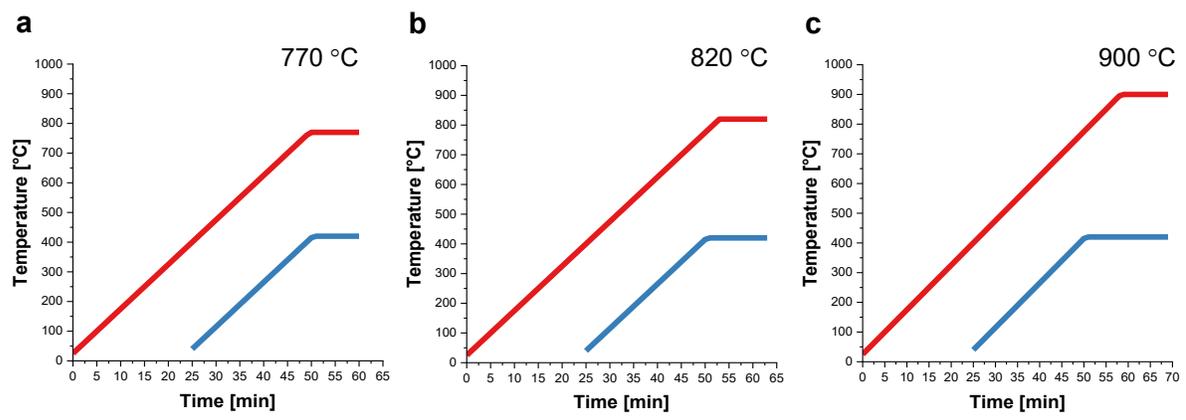

**Figure S3:** CVD temperature profiles for 770 °C (a). 820 °C (b) 900 °C (c). The red curves represent the heating profiles of the transition metal precursors, while the blue curves correspond to the heating of the selenium precursor.

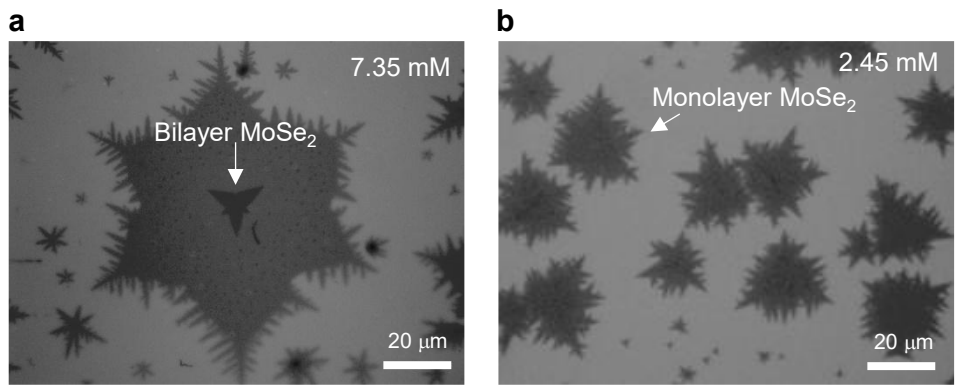

**Figure S4:** Optical microscopy image of bilayer (2L) (a) and monolayer (b) MoSe$_2$ grown using 7.35 mM and 2.45 mM of Na$_2$MoO$_4$ aqueous solution, respectively.

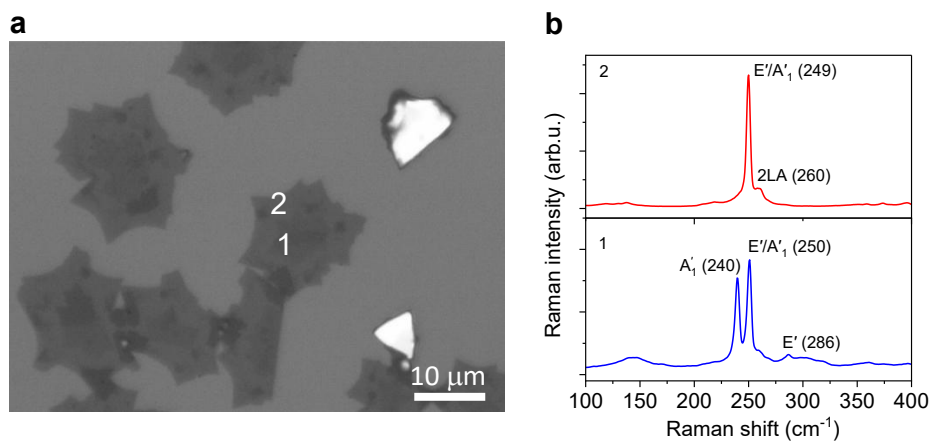

**Figure S5**: (a) Optical microscopy image of a mixed-phase lateral HS, and (b) corresponding Raman spectra collected from the inner region (1) and outer region (2). Region 1 contains both $MoSe_2$ and $WSe_2$ phases, while Region 2 shows only the $WSe_2$ phase.

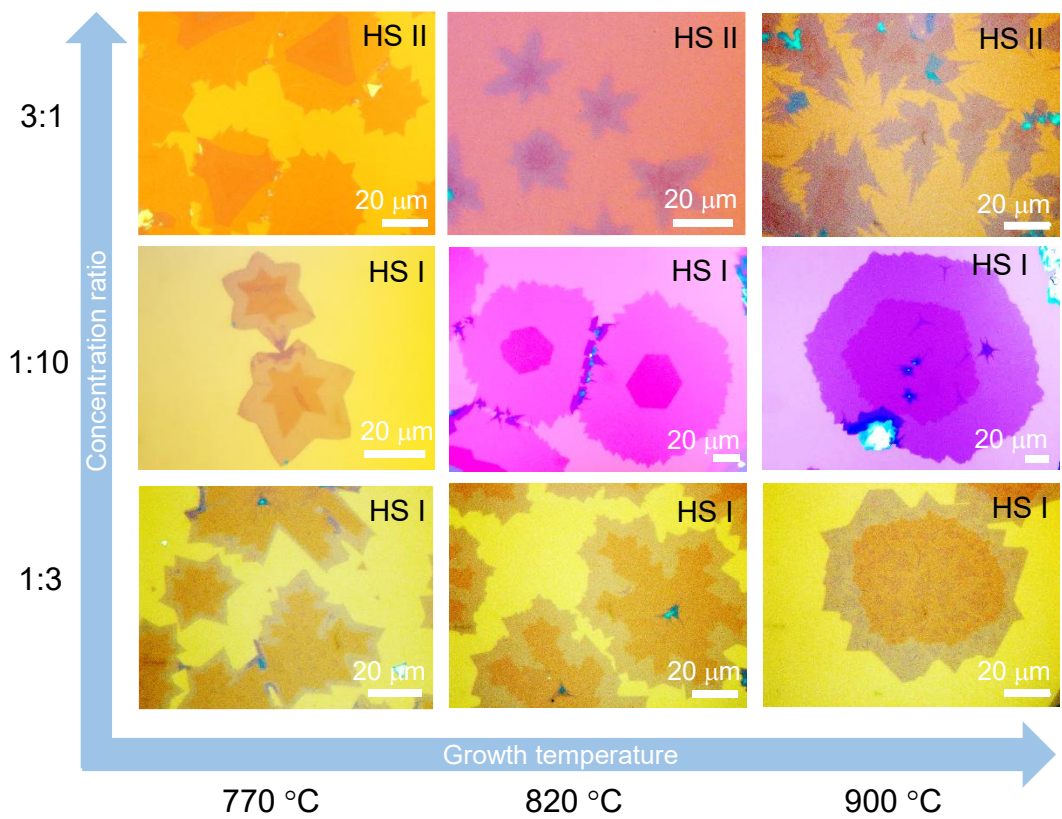

**Figure S6**: False color optical microscopy images of HS I and HS II as a function of growth temperature and precursor concentration ratio, showing the dependence of crystal size and morphology on synthesis conditions. Detailed analysis of crystal size is demonstrated in **Figure S7.**

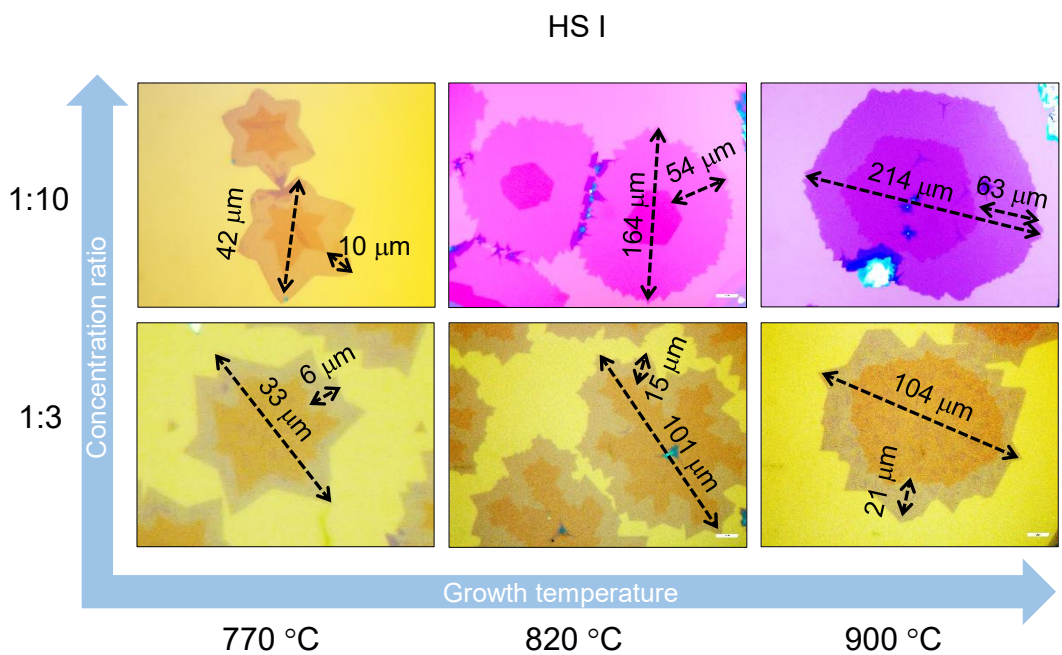

**Figure S7**: False color optical images of HS I as a function of growth temperature and precursor concentration ratio, showing the dependence of overall crystal size and width of WSe$_2$ on synthesis conditions. The scale bar is identical to that in **Figure S6**.

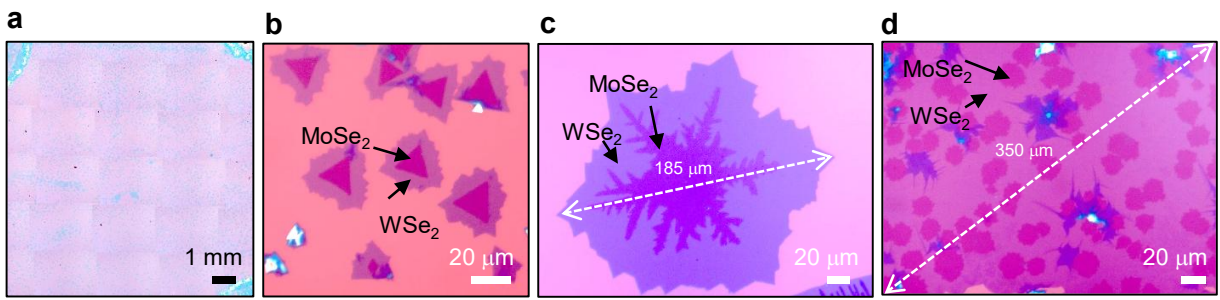

**Figure S8**: (a) HS I is grown uniformly across the entire substrate (12 × 12 mm²). demonstrating the scalability of HS I growth using liquid precursors. Enlarged image (b) of isolated HS I (~10-50 µm) at the center (c) big isolated HS I (~185 µm) closer to the edge (d) continuous HS I (~350 µm) at the edge of the growth substrate.

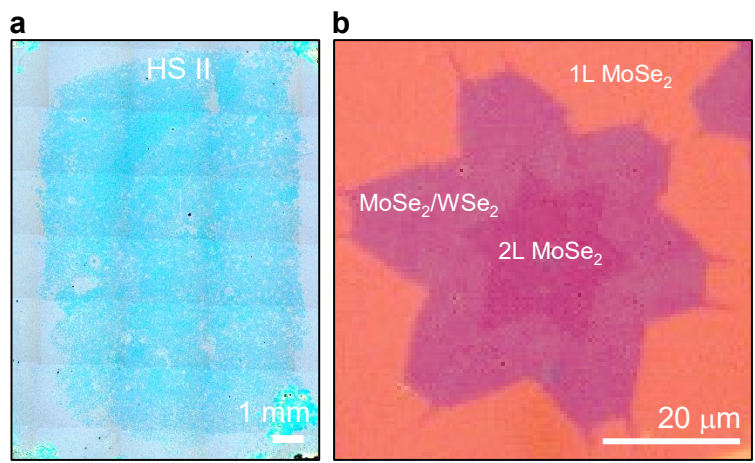

**Figure S9**: (a) HS II is grown uniformly across the substrate (8 × 12 mm$^2$), demonstrating the scalability of HS II growth using liquid precursors. (b) enlarged image of a HS II. Note that the blocky appearance in (a) is a stitching artifact.

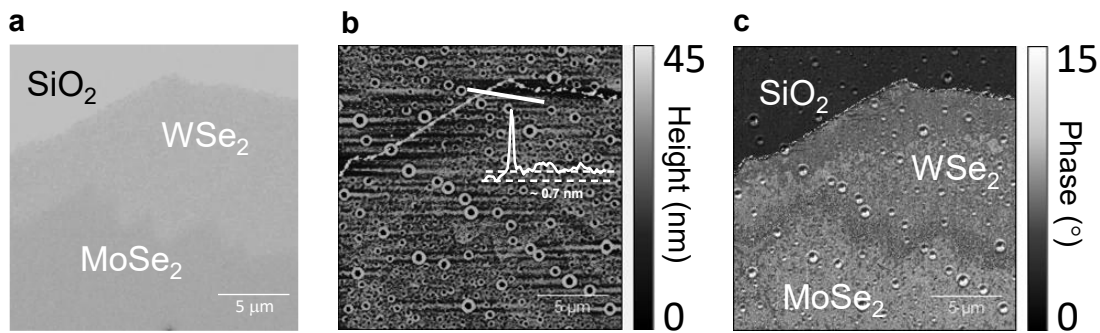

**Figure S10**: Microscopic characterization of HS I. (a) Optical microscopy image of MoSe$_2$-WSe$_2$ HS I. (b) The corresponding topography AFM image. The inset shows the height profile along the white line. (c) shows the phase image. The phase image identifies the interface between the MoSe$_2$ and WSe$_2$ regions. Additionally, in the AFM images, the presence of black circular spots across the lateral HS and throughout the substrate can be recognised. This likely arises from precursor segregation (Figure S11), and the formation of Na$_2$O as a byproduct during annealing[1-3], which can react with SiO$_2$ to form Na$_4$SiO$_4$ [4], leading to surface etching[5]. Despite this, Raman mapping (Figure 2b, c) and TEM analyses (Figure S12, 3a) confirm a continuous, high-quality MoSe$_2$-WSe$_2$ lateral HS.

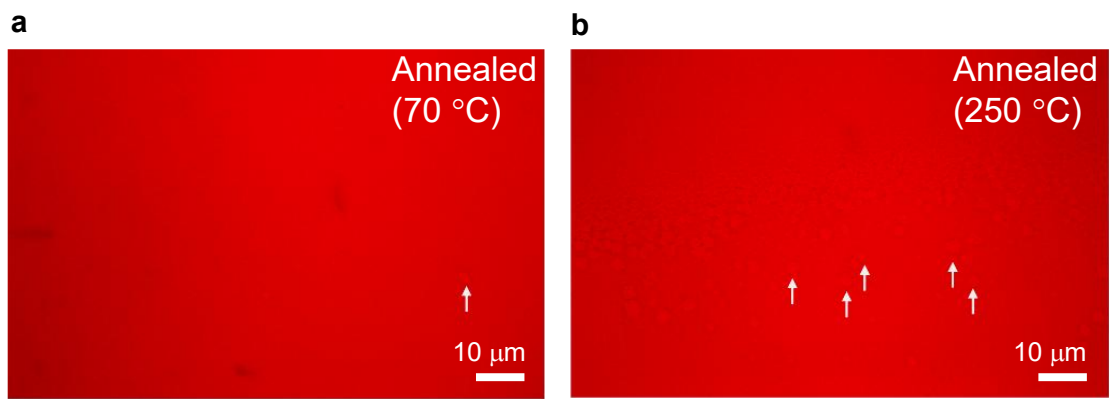

**Figure S11:** Optical microscopy images after heating of deposited liquid precursor which forms circular patterns. (a) at 70 °C and (b) 250 °C.

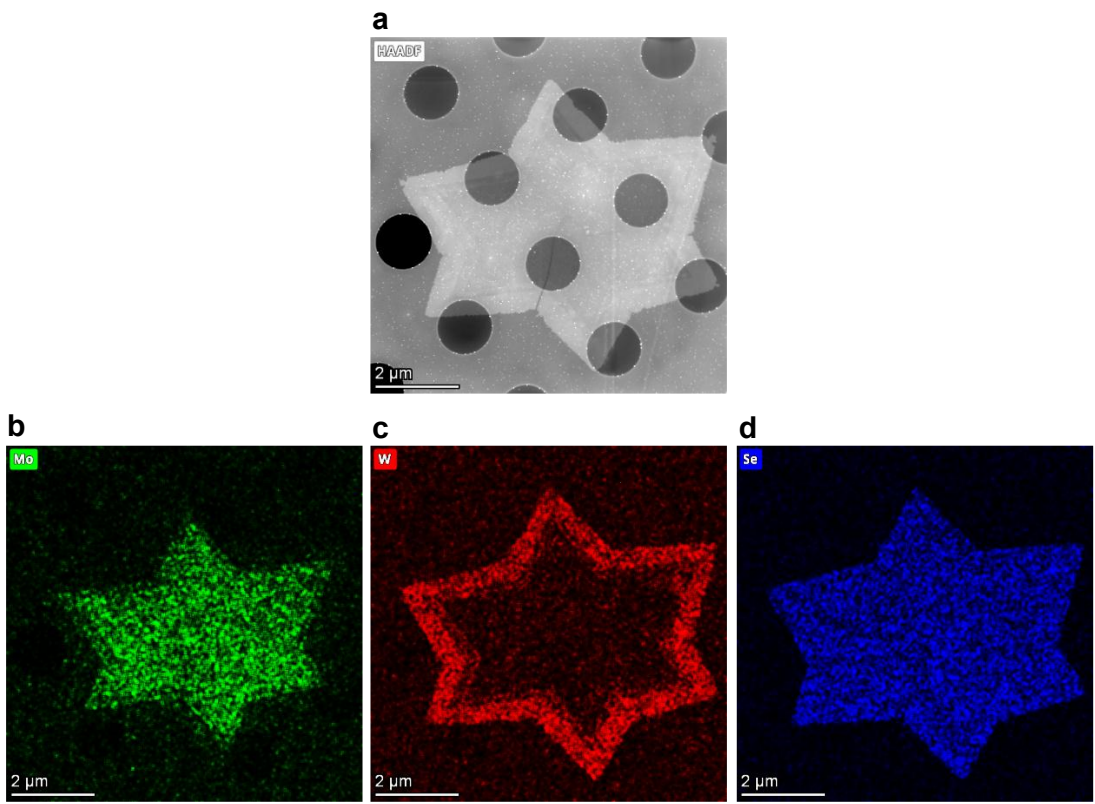

**Figure S12:** (a) 200 kV-HAADF-STEM image of the heterostructure. (b-d) EDX elemental maps representing the spatial distribution of Se (blue), Mo (green), and W (red).

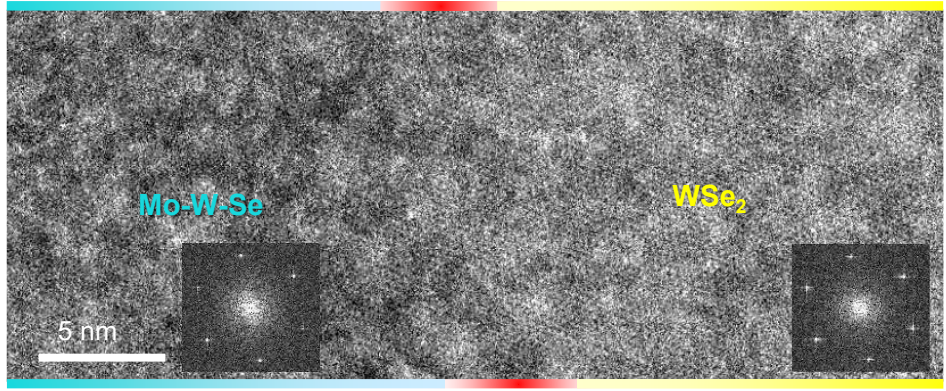

**Figure S13:** (a) High-magnification 200 kV HAADF-STEM image of the Mo-W-Se and WSe$_2$ boundary shows a narrow ≈ 5 nm interface (red lines), with Mo-W-Se and WSe$_2$ regions indicated by cyan and yellow lines, respectively. The inset displays the corresponding fast Fourier transforms (FFTs) obtained from the respective regions.

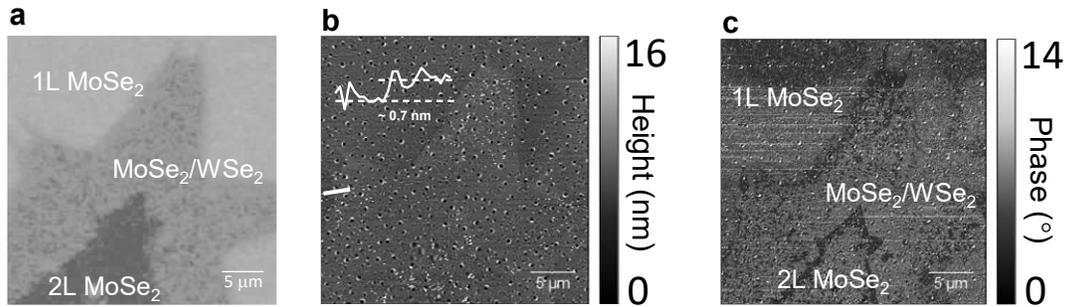

**Figure S14:** Microscopic characterization of HS II. (a) Optical microscopy image of MoSe$_2$-WSe$_2$ HS II. (b) represents an AFM topography image. Inset shows the height profile along the white line. (c) shows the phase image, which highlights the three phases -1L MoSe$_2$, MoSe$_2$/WSe$_2$ heterobilayer, and homobilayer (2L) MoSe$_2$. In the phase image, two interfaces between the bottom MoSe$_2$ and the edge of the top MoSe$_2$-WSe$_2$ layer, and between WSe$_2$ and MoSe$_2$ regions in the top MoSe$_2$-WSe$_2$ layer can be recognized. The phase and height change can be found at the interface of the bottom MoSe$_2$ and edges of the MoSe$_2$-WSe$_2$ layer, however, only the phase change and no height change are noticeable at the interface of top WSe$_2$ and top MoSe$_2$, which further confirms the formation of the expected ML lateral HS on the bottom MoSe$_2$ layer.

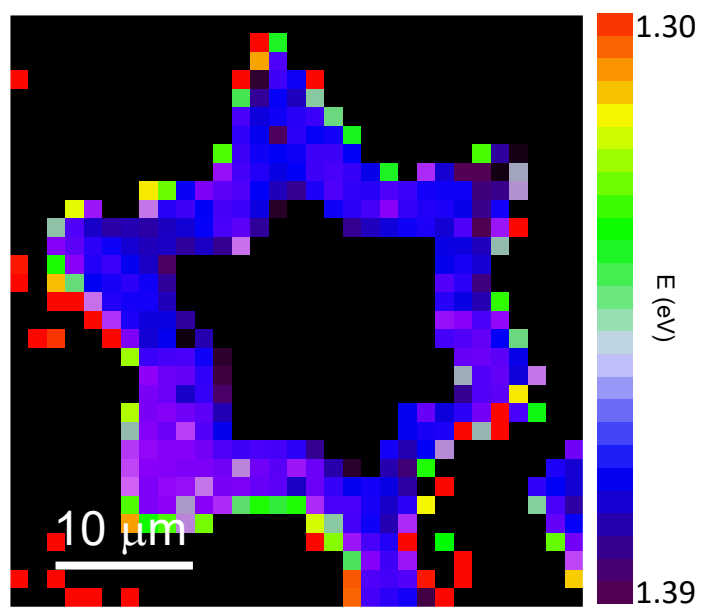

**Figure S15.** Representative interlayer exciton (IX) emission map at room temperature of $MoSe_2/WSe_2$ heterobilayer in HS II.

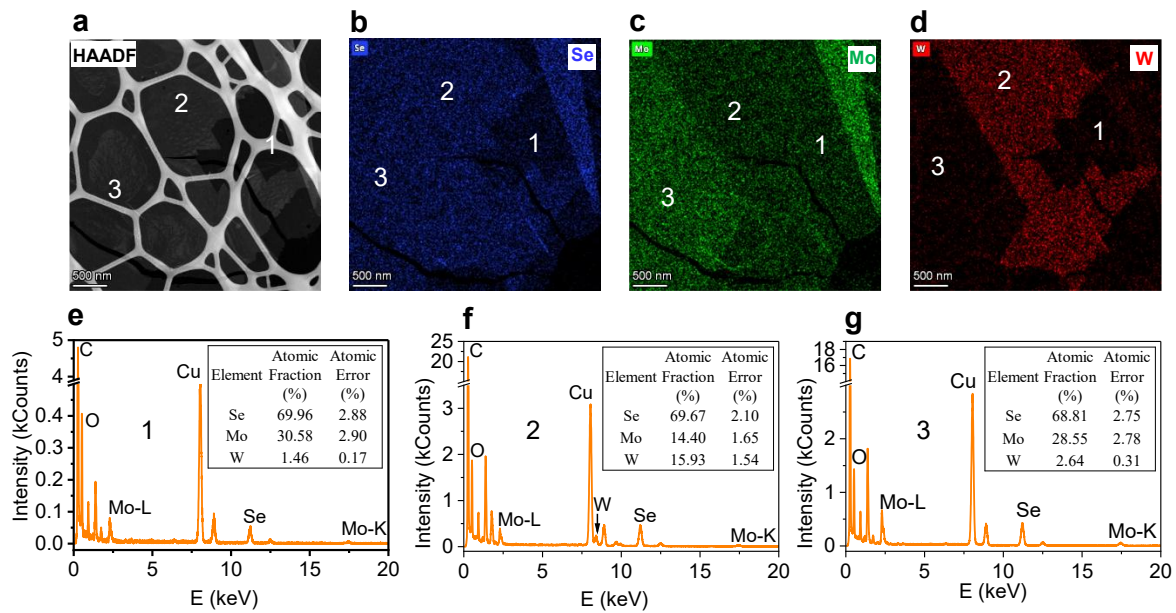

**Figure S16:** (a) 80 kV HAADF-STEM image of a HS II. (b-c) EDX elemental maps representing spatial distribution of Se (blue), Mo (green), and W (red). (e-g) Corresponding EDS spectra, with insets showing the atomic percentages of Se, Mo, and W at regions 1 (1L $MoSe_2$), 2 ($MoSe_2$/$WSe_2$ heterobilayer), and 3 (2L $MoSe_2$), respectively.

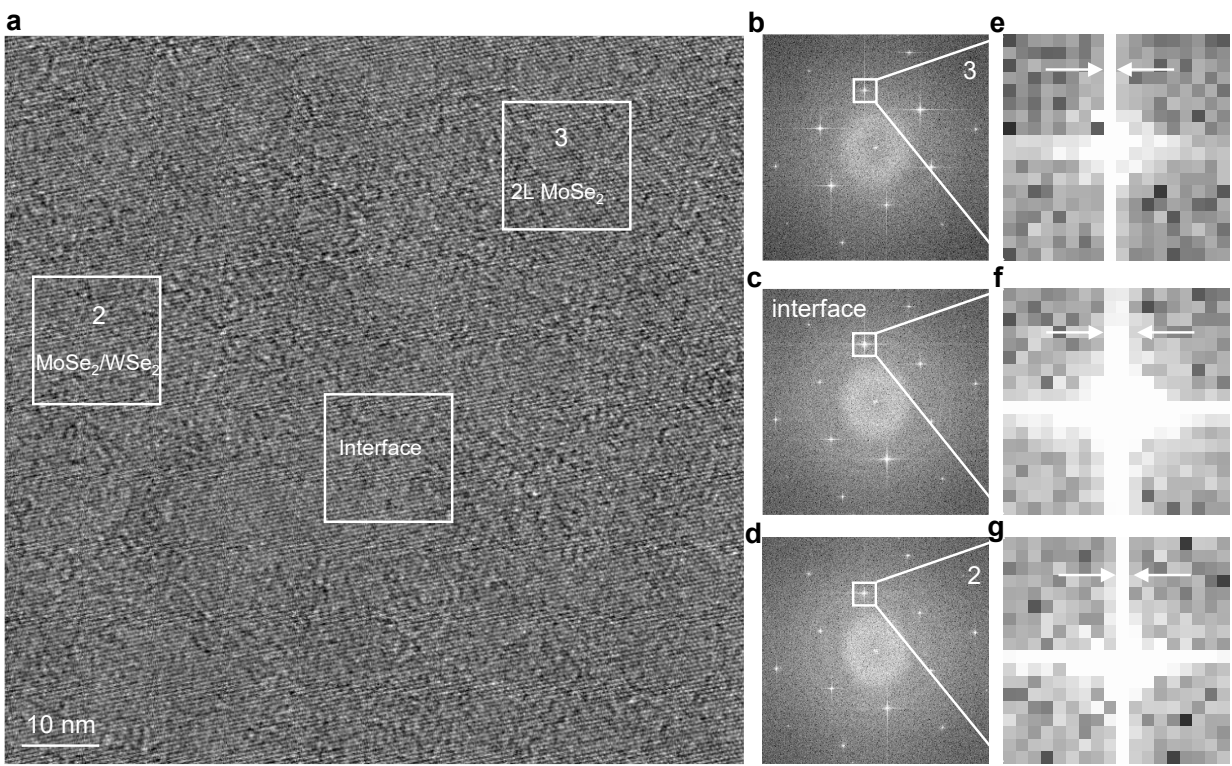

**Figure S17:** (a) 200 kV High-resolution TEM (HRTEM) image of the top lateral heterostructure in HS II showing continuous atomic rows across the top layer, indicative of epitaxial growth of the top LH. (b-d) Fast Fourier transform (FFT) patterns obtained from three different regions ($MoSe_2$ homobilayer, interface of top LH and $MoSe_2/WSe_2$ heterobilayer). (e-g) Corresponding magnified images of single diffraction spots highlight the transition from the homobilayer $MoSe_2$ (region 3) to the heterobilayer $MoSe_2/WSe_2$ (region 2). While regions 2 and 3 exhibit narrower fringes, the interface displays a wider fringe. The fringe shift between regions 2 and 3 supports the transitional boundary from homobilayer to heterobilayer. Please note that remaining amorphous and graphitic surface contaminations of residual hydrocarbons as results of the transfer process are overlaying the lattice structure of the TMDs and lowering the quality of the HRTEM images.

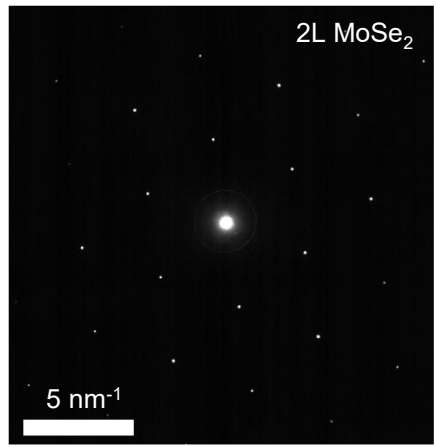

**Figure S18:** Selected-area electron diffraction (SAED) patterns acquired at 200 kV of bilayer (2L) MoSe$_2$ region of HS II display six-fold symmetry, characteristic of the hexagonal lattice. The presence of hexagonal diffraction pattern with circular spots supports the epitaxial growth of the top MoSe$_2$ ML.

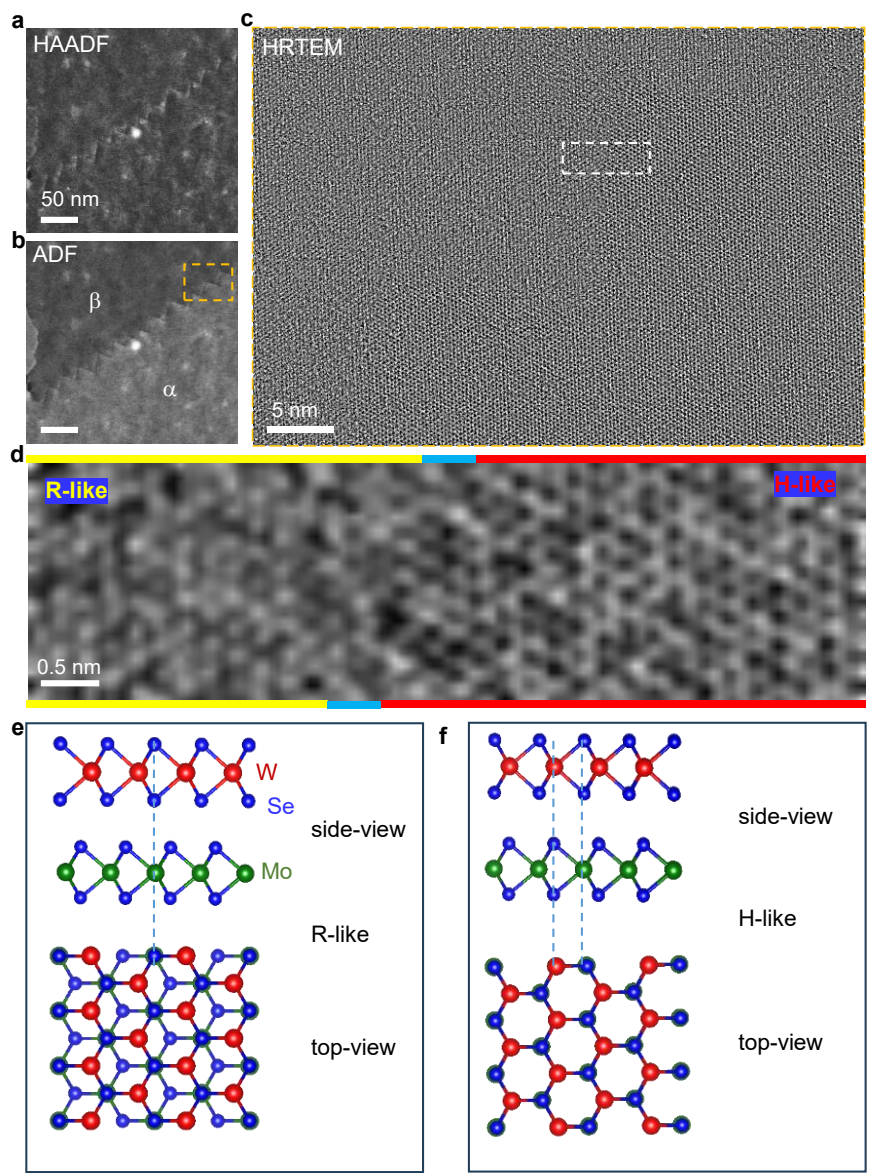

**Figure S19:** MoSe$_2$/WSe$_2$ heterobilayer with H and R-like stacking. (a,b) Comparison of 200 lV HAADF- and ADF-STEM images highlighting two domains (α and β, also denoted in main *figure 5h*), which are clearly resolved by brighter and darker contrast in the ADF image. (c) 200 kV high-resolution TEM image of the region marked by the orange dashed box in (b). (d) Magnified view of the area outlined by the white dashed box in (c), showing regions with AA (R-like) and AB (H-like) stacking. (e,f) Corresponding atomic models illustrating the stacking arrangements. In the AA (R-like) stacking, Se atoms of WSe$_2$ align above Mo atoms of MoSe$_2$, whereas in the AB (H-like) stacking, W (Se) atoms of WSe$_2$ are positioned above Se (Mo) atoms of the MoSe$_2$ layer, thereby forming a hollow hexagonal structure when viewed from the c-axis direction (top view).

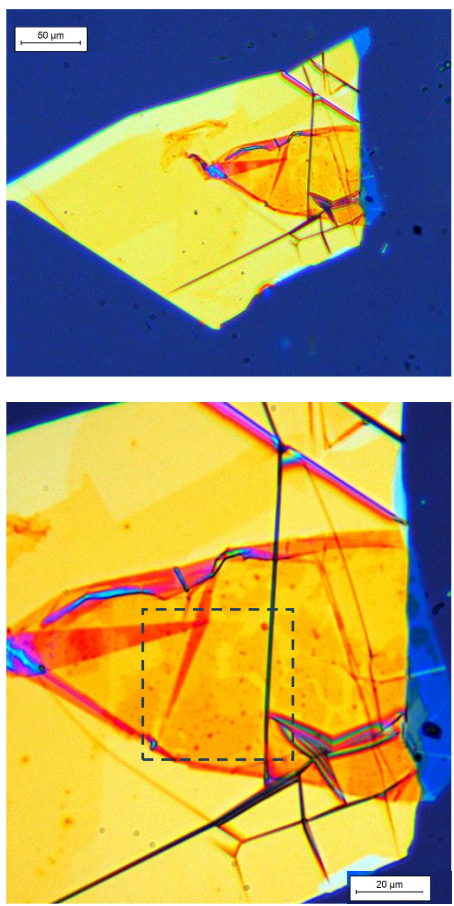

**Figure S20:** Optical microscope images of the hBN encapsulated HS II.

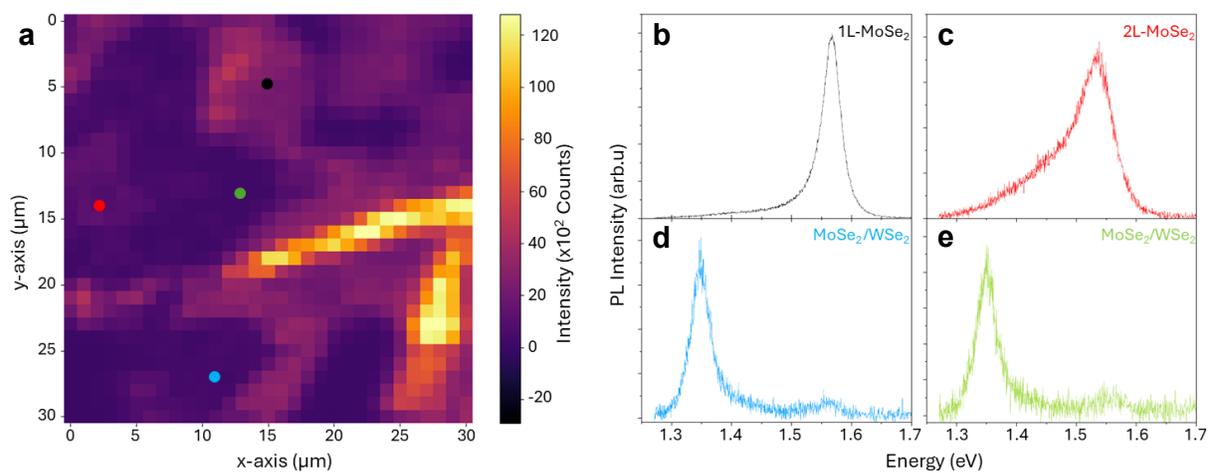

**Figure S21:** Room temperature (T = 300K) PL spectral map of hBN encapsulated HS II. (a) PL map of encapsulated HS II. Spectra corresponding to the regions marked with colored dots are presented in b-e. PL spectra from (b) 1L-MoSe$_2$ (c) 2L-MoSe$_2$ and (d-e) MoSe$_2$-WSe$_2$ heterobilayer regions.

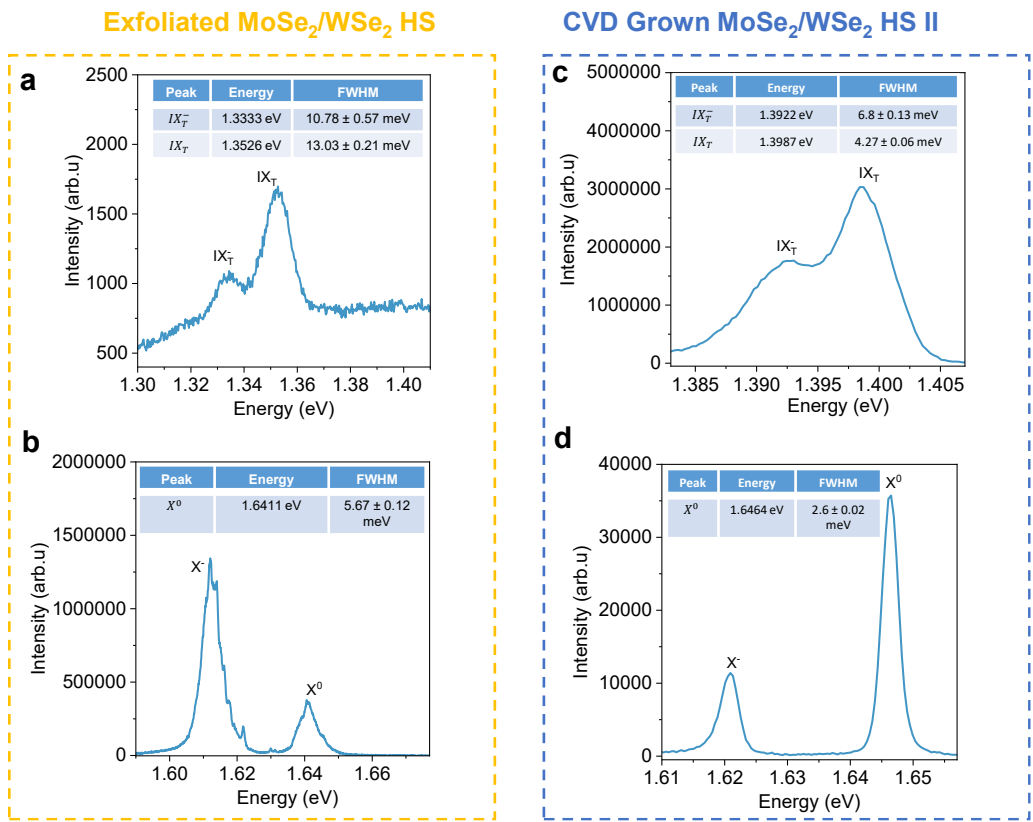

**Figure S22:** Comparison of PL spectra of the exfoliated MoSe$_2$/WSe$_2$ vertical HS and the CVD-grown MoSe$_2$-WSe$_2$ HS II at 4K.

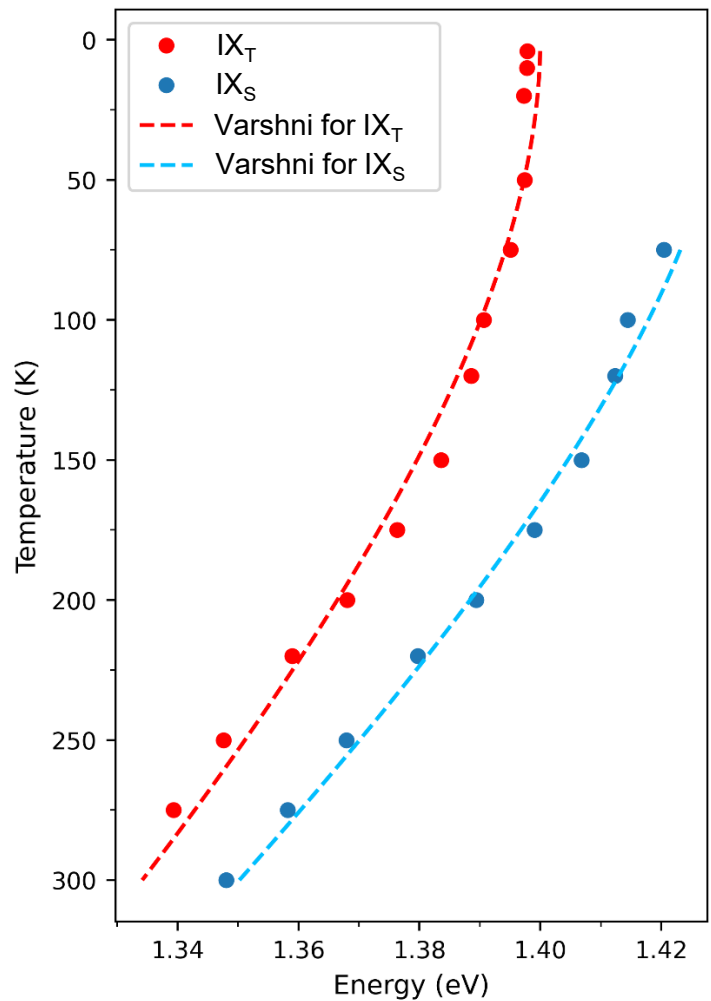

**Figure S23:** Varshni function fits for the temperature dependent peak shifts of $IX_T$ (in red) and $IX_s$ (in blue).

**Table S1**: Ratio and molar concentrations of transition metal precursors in aqueous solution.

| Ratio C1/C2 | $Na_2MoO_4$ [mM] | $Na_2WO_4$ [mM] |
|---|---|---|
| 1:3 | 2.45 | 7.35 |
| 3:1 | 7.35 | 2.45 |
| 1:10 | 2.45 | 24.5 |

**Table S2:** Varshini Parameters for 1L-MoSe$_2$, IX$_T$, IX$_s$ from the HS II sample and from exfoliated 1L-WSe$_2$ for reference.

| Sample | E$_0$ [eV] | α [eV/K] | β [K] |
|---|---|---|---|
| MoSe$_2$ (HS II) | 1.64 | 9.3E-4 | 754 |
| WSe$_2$ (Exfoliated) | 1.72 | 6.0E-4 | 331 |
| IX$_T$ (HS II) | 1.40 | 5.7E-4 | 479 |
| IX$_s$ (HS II) | 1.43 | 5.7E-4 | 335 |